\documentclass[11pt,letterpaper]{article}
\usepackage[margin=2.5cm]{geometry}
\usepackage{amssymb,amsmath,epsfig,esint}
\usepackage{CJKutf8}
\usepackage{amsfonts}
\usepackage{setspace,lipsum}
\usepackage{graphicx}
\usepackage{subcaption}
\usepackage[english]{babel}
\usepackage{booktabs}
\usepackage{cellspace}
\usepackage{multicol}
\usepackage{multirow}
\usepackage{float,tabularx,siunitx,array}
\usepackage[utf8]{inputenc}
\usepackage{authblk}
\usepackage{fourier} 
\usepackage{array}
\usepackage{makecell}
\usepackage[toc,page]{appendix}
\usepackage[nottoc]{tocbibind}
\usepackage[colorlinks]{hyperref}
\usepackage{url}
\hypersetup{
    colorlinks=true,
    linkcolor=red,
    filecolor=magenta, 
    citecolor=blue,        % color of links to bibliography
    urlcolor=blue,
    pdftitle={},
    pdfpagemode=FullScreen,
    }

\numberwithin{equation}{section}
\newcolumntype{C}[1]{>{\centering\arraybackslash}p{#1}}

\newcommand{\eref}[1]{Eq.~(\ref{#1})}

\usepackage[explicit]{titlesec}
\titleformat{\section}
{\normalfont\normalsize\bfseries}{\thesection}{12pt}{\filcenter
\MakeUppercase{#1}} %to get the section heading in center, the deired size and Upper casing (capital)
\titleformat{\subsection}
{\normalfont\normalsize\bfseries}{\thesubsection}{12pt}{\filcenter{#1}} %#1 refers to the subsection title in the text

\begin{document}
\title{\bf Dynamics of interacting scalar field model in the realm of\\ chiral cosmology }
\author{Trupti Patil\thanks{trupti19@iiserb.ac.in}\hspace{0.1cm},\hspace{0.1cm} Sukanta Panda
\thanks{sukanta@iiserb.ac.in}\hspace{0.1cm}, \hspace{0.1cm} Manabendra Sharma$^\diamond$  \thanks{sharma.manabendra1501@yahoo.com}\hspace{0.2cm} and\hspace{0.1cm} Ruchika\thanks{ruchika@ctp-jamia.res.in}  \\
$^\ast$$^\dagger$Department of Physics, IISER Bhopal,\\
Bhopal - 462066, India\\
$^\diamond$Inter University Centre for Astronomy $\&$ Astrophysics, Post Bag 4, Pune-411007, India.\\
$^\ddagger$NAS, Centre for Theoretical Physics $\&$ Natural Philosophy, Mahidol University,
Nakhonsawan Campus, Nakhonsawan 60130, Thailand.\\
$^\S$Department of Physics, Indian Institute of Technology, Bombay, Mumbai-400076 India
}

\date{}
\maketitle
\begin{abstract}
The strange behaviour of the universe's dark sector offers us the flexibility to address cosmological problems with different approaches. Using this flexibility, we consider a
possible exchange of energy among the dark sector components as a viable candidate model. 
In the present work, we investigate the interaction between two scalar fields within the generalization of a two-component chiral cosmology. We also show that there exists a unique equivalence between fields and fluids description of interacting dark sector model. Later, a detailed analysis of the dynamics of the dark energy-dark matter model with coupling in both kinetic and potential parts has been performed using a method of qualitative analysis of dynamical systems. Moreover, the cosmological viability of this model is analyzed for the potential of an exponential form via the phase-space study of autonomous system for various
cosmological parameters.
\end{abstract}

\section{Introduction}\label{1}
After the discovery of the accelerated expansion of the universe at the very end of the twentieth century, various theoretical models have been recommended to comply with the experimentally obtained evidences\cite{1998, Perlmutter_1999, Riess_1998, 1999, Dunkley_2009, Boughn:2004zm, 2005, PhysRevD.69.103501, 2007, PhysRevD.74.123507,DiValentino:2019jae} and to set up a new context for the physics of the future. In the quest to explain recent cosmological observations, a novel concept of dark energy with negative pressure has been introduced and studied closely in the literature \cite{2006, article}. The mysterious nature of this dark entity has multiplied the area of research with a focus on uncovering its properties and it has been one of the hot topics in modern cosmology. Among the various choices for dark energy (DE), the cosmological constant is the one that favours the observations most \cite{2020}. Nevertheless, as seen from the theoretical point of view, it consists of two fundamental problems in cosmology, 1) cosmological constant problem \cite{RevModPhys.61.1, Doran:2002ec} and 2) coincidence problem \cite{Wang_2016,PhysRevLett.82.896, 2014, 2009, 2016}. 
\par
To overcome these problems various other canonical \cite{potting2021coupled, Harko2014ArbitrarySA} and non-canonical scalar field models \cite{Mandal:2021ekc, 2021} were introduced. It has been suggested and studied that dark energy could be dynamic and evolve with time. This feature has been studied in model where equation of state parameter $\omega_{\phi}$ ranges as $-1<\omega_{\phi}<-\frac{1}{3}$ \cite{potting2021coupled, Harko2014ArbitrarySA, GonzlezDaz2000CosmologicalMF,refId0} and leads to beat the cosmological constant problem. The various other dark energy scalar field model viz. k-essence, phantom, quintom, tachyon, galileon, and multi-scalar field models \cite{PhysRevD.63.103510,Sen:2004nf,Bamba:2012cp,VIJAYASANTHI2022101725,1995,FENG200535,PhysRevD.71.063513, Anto, ROZASFERNANDEZ2012313,2008mgm..conf.1761D}, where the non-canonical kinetic term appears as a coupling factor have also been studied meticulously and found to be very useful to ease the coincidence problem. However, there are recent studies that argue a large class of Dark Energy Models, including Quintessence and K-essence \cite{salv, Lee_2022, Banerjee,Oikonomou:2019muq}. More often than not we address those models as dark energy - dark matter interaction models. Various DE interacting scalar field models \cite{potting2021coupled,2021,2017,chakraborty2020dynamical,Guo_2005,Guo2005CosmologicalEO,Chakraborty:2020vkp} and phenomenological fluid models \cite{PhysRevD.104.063517,Caldera-Cabral:2008yyo,10.1093/mnras/stx2278,10.1093/mnras/stv1495,Odintsov:2017icc,Oikonomou:2019boy,Odintsov:2018uaw} have been set forth and studied in literature. Another approach to address this issue is based on modified gravity (See Ref. \cite{Rudra:2014xba,Vinutha:2021jbs,Clifton:2011jh} and references therein).
\par
The dynamics of chiral cosmology has been studied earlier in \cite{Paliathanasis:2018vru,Paliathanasis_2020,Chervon2013ChiralCM,Paliathanasis:2020abu}. It has been noticed that chiral cosmology can provide a suitable realm for the description of the multi-scalar field model and draw the framework where characteristics of the dark sector universe can be explained under the generalization of the chiral cosmological model. \par
In our work, we are interested in the background dynamics of field equations in general relativity in a spatial flat Friedmann-Lema$\hat{i}$tre-Robertson-Walker (FLRW) (homogeneous and isotropic) universe. Motivated by quintom paradigm \cite{Cai:2009zp}, in our analysis, we aim to study a two-field interaction model under the chiral cosmological framework. We consider a case where the coupling factor lives in kinetic as well in potential parts of the fields. The findings, e.g. in \cite{Nozari:2016ilx} and the study of \cite{Lazkoz:2006pa} show that such DDI models are indeed more capable than was known up till that can solve current recurring problems in cosmology, e.g. coincidence problem and thus, in our opinion, makes it an interesting study to probe further.\par
The equivalence between the phenomenological fluid model and the 2-field scalar field model provides us with the interaction term which is a mixture of the dark energy field and dark matter fluid energy density \cite{PhysRevD.103.023510}. In such a scenario, the rigorous form of the Einstein field equations is turned into an autonomous system of equations by defining preferable dimensionless variables. In particular, we have found that the simple dynamical system evaluation \cite{article,BAHAMONDE20181} of an interacting DE-DM model issues the physical validity of the discussed cosmological DE model in late-time cosmology.  The theory can be applicable to address the current cosmological problems as long as the background dynamics are studied, where we notice sufficiently long, extended era of matter dominated universe to the current era of an accelerated expanding universe. Additionally, we have examined that the big rip singularity problem does not appear in our model. The plan of the work is structured as follows: In Section \ref{2}, we describe the chiral cosmology and set up dynamical equations for the two-field chiral model. In Section \ref{3}, we consider a coupled dark energy (DE) - dark matter (DM) model and obtain evolution equations. Section \ref{4} systematically shows a one-to-one correspondence between the fluid and field theory approaches. The principal analysis of our work is presented in Section \ref{5}. It explains the formulation of an autonomous system consisting of non-linear coupled differential equations. The examination of critical points of the system and their stability analysis have also been shown in this section. In the subsection, we show the graphical representation of the obtained results to complement the physical viability of the model. The final Section \ref{6}, summarizes the present work and outlines the conclusion.

\section{ Chiral cosmology} \label{2}
The gravitational action integral for the chiral cosmological model reads as:
\begin{equation}
\label{eqn:1}
    S = \int d^4x \sqrt{-g}\bigg(\frac{R}{2} - \frac{1}{2} g^{\mu \nu} F_{AB}(\varphi) \nabla_{\mu} \varphi^A \nabla_{\nu} \varphi^B - V(\varphi)\bigg). 
\end{equation} 
where, R is the Ricci scalar. $g_{\mu \nu}$ is the metric tensor of the 4-D space-time along with $ F_{ab}(\varphi)$ being the second-rank tensor of chiral space where the scalar field evolve. $ V(\varphi)$ is the potential. Here, we use the natural units where c = 1 and  the reduced Planck mass $M_p^2 = \frac{1}{8 \pi G}=1$.\\[0.2cm]
The individual variation of Eq. \eqref{eqn:1}  w.r.t. the metric tensor and the scalar fields $\varphi^A$ brings us the Einstein's gravitational field equations and the Klein-Gordon equation as:
\begin{equation}
\label{eqn:2}
    G_{\mu \nu} = T_{\mu \nu} \equiv F_{AB}(\varphi) \nabla_{\mu} \varphi^A \nabla_{\nu} \varphi^B - g_{\mu \nu} \bigg(\frac{1}{2} g^{\mu \nu} F_{AB}(\varphi) \nabla_{\mu} \varphi^A \nabla_{\nu} \varphi^B + V(\varphi)\bigg) ,
\end{equation}
\begin{equation}
\label{eqn:3}
    g^{\mu \nu} \bigg(\nabla_\mu F^A _{B}(\varphi) \nabla_{\nu} \varphi^B\bigg) - F^A _{B}(\varphi)  \frac{\partial V(\varphi)}{\partial  \varphi^B} =0. 
\end{equation}
The line element for the chiral space is given as:
\begin{equation}
\label{eqn:4}
    ds^2_c = F_{AB}(\varphi) d\varphi^A d\varphi^B, \hspace{0.5cm}
    A,B=1,2,...,N .
\end{equation}
\vspace{0.1cm}
\subsection{2-field chiral model}
To study the interaction between the dark sector fields, we consider a two scalar field chiral cosmological model. A possible interaction in both kinetic and potential parts can be seen between the two fields \cite{PhysRevD.103.023510,Paliathanasis:2020sfe} which can be an effective mechanism to explain the coincidence problem and cosmological constant problem. For further analysis of the model we set,\\

$\varphi^1 = \phi$ \hspace{0.2cm} and\hspace{0.2cm} $\varphi^2 = \psi$  \hspace{0.2cm}and \\

$F_{11}= A(\phi, \psi)$, \hspace{0.2cm} $F_{22}= B(\phi, \psi)$, \hspace{0.2cm} $F_{12}= C(\phi, \psi)=F_{21}$.\\
\newline
Following the analysis in the chiral space, for an isotropic and homogeneous universe, the space-time \eqref{eqn:4} gives us the spatially flat FLRW space-time with the line element  as
\begin{equation}
\label{eqn:9}
    ds^2 = -dt^2 + a^2(t) \big(dx^2+dy^2+dz^2 \big).
\end{equation}
where a(t) defines the scale factor as a function of cosmic time.\\
The gravitational action integral \eqref{eqn:1} in the case of two scalar fields then becomes:
\begin{multline}
\label{eqn:5}
     S = \int d^4x \sqrt{-g}\bigg( \frac{R}{2} - \frac{1}{2} g^{\mu \nu}A(\phi, \psi) \phi,_{\mu} \phi,_{\nu} - \frac{1}{2} g^{\mu \nu}B(\phi, \psi) \psi,_{\mu} \psi,_{\nu}
     - g^{\mu \nu}C(\phi, \psi) \phi,_{\mu} \psi,_{\nu}
     - V(\phi, \psi)\bigg).
\end{multline}\\[0.2cm]
Also, Eq. \eqref{eqn:2} provides us the Einstein's gravitational field equation as:
\begin{multline}
\label{eqn:6}
    G_{\mu \nu} = T_{\mu \nu} \equiv A \phi,_{\mu} \phi,_{\nu} + B \psi,_{\mu} \psi,_{\nu} + 2C \phi,_{\mu} \psi,_{\nu} 
    - g_{\mu \nu} \bigg(\frac{1}{2} g^{\mu \nu} (A \phi,_{\mu} \phi,_{\nu} + B \psi,_{\mu} \psi,_{\nu} + 2C \phi,_{\mu} \psi,_{\nu}) + V(\phi, \psi)\bigg).
\end{multline}
Furthermore, Eq. \eqref{eqn:3} provides the Klein-Gordon equations in the FRW universe as:
\begin{eqnarray}
3H(A \dot{\phi} + C \dot{\psi}) + \partial_t (A \dot{\phi} + C \dot{\psi}) - \frac{1}{2} \frac{\partial A}{\partial \phi} \dot{\phi}^2 - \frac{\partial C}{\partial \phi} \dot{\phi} \dot{\psi} - \frac{1}{2} \frac{\partial B}{\partial \phi} \dot{\psi}^2 + \frac{\partial V}{\partial \phi} = 0 ,  \label{eq:7}\\[0.2cm]
3H(C \dot{\phi} + B \dot{\psi}) + \partial_t (C \dot{\phi} + B \dot{\psi}) - \frac{1}{2} \frac{\partial A}{\partial \psi} \dot{\phi}^2 - \frac{\partial C}{\partial \psi} \dot{\phi} \dot{\psi} - \frac{1}{2} \frac{\partial B}{\partial \psi} \dot{\psi}^2 + \frac{\partial V}{\partial \psi} = 0 .  \label{eq:8}
\end{eqnarray}
where, $H=\frac{\dot{a}}{a}$ is the Hubble rate.
\par
In the next section, we consider a two scalar fields interacting model where both the fields are coupled to each other. The coupling can be observed either in kinetic or in potential or in both parts of the setup Lagrangian. This kind of model has been considered before as well in literature \cite{FENG200535,Chervon2013ChiralCM,Diaz-Barron:2021ari}. The objective behind studying such a model is to achieve the negative running of the equation of state (EoS) parameter which is also consistent with present observational scenarios.

\section{Construction of DE-DM interaction model}  \label{3}
As inspired by chiral cosmology \cite{Paliathanasis:2018vru,Paliathanasis_2020,Chervon2013ChiralCM,Paliathanasis:2020abu}, we work with a model where the parameters in Eq. \eqref{eqn:5} take value as:
\begin{center}
    A = -1,   \hspace{0.2cm}   B = B($\phi$),  \hspace{0.2cm}  C = 0,
\end{center}
for which action integral is given as:
\begin{equation}
\label{eqn:10}
    S =  \int d^4x \sqrt{-g}\bigg( \frac{R}{2} + \frac{1}{2} g^{\mu \nu} \phi,_{\mu} \phi,_{\nu} - \frac{1}{2} B(\phi) g^{\mu \nu} \psi,_{\mu} \psi,_{\nu} - V(\phi) - \alpha \phi^2 \psi^2\bigg).
\end{equation}
One of our prime goals is to work on assimilating DE-DM interaction in action \eqref{eqn:10}. In order to accomplish our goal to design a DE-DM interaction (DDI) model, we here, consider one of the fields as a dark energy candidate and another one as a dark matter candidate. %(Please note that we do not limit any particular field to play a particular role of dark sector components at this time. The details about this have been studied in the following section.)
The details about the roles played by the fields have been discussed in the following section.
The last term in Eq. \eqref{eqn:10} represents an interacting potential.
\\
Now, from the line element \eqref{eqn:9} and modified action \eqref{eqn:10}, the dynamical field equations for the fields $\phi$ and  $\psi$ are given, respectively, as in \eref{eq:11} and \eref{eq:12}
\begin{eqnarray}
\nabla_{\mu} \nabla^{\mu} \phi + \frac{B,_\phi}{2} \nabla_{\mu} \psi \nabla^{\mu} \psi + 2\alpha \phi \psi^2 + V,_\phi = 0, \label{eq:11} \\
\nabla_{\mu} \nabla^{\mu} \psi + \frac{B,_\phi}{B} \nabla_{\mu} \psi \nabla^{\mu} \phi - 2\alpha \psi \phi^2 B^{-1} = 0.  \label{eq:12}
\end{eqnarray}
where, \hspace{0.1cm}`$,_{\phi}$' denotes the partial derivative w.r.t $\phi$.\\
Whereas the Friedmann equations are given as in \eref{eq:13} and \eref{eq:14}
\begin{eqnarray}
3H^2 = \frac{-\dot{\phi}^2}{2} + \frac{B(\phi)}{2}\dot{\psi}^2 + V(\phi) + \alpha \phi^2 \psi^2 , \label{eq:13} \\
2\dot{H} + 3H^2 =-\bigg( \frac{-\dot{\phi}^2}{2} + \frac{B(\phi)}{2}\dot{\psi}^2 - V(\phi) - \alpha \phi^2 \psi^2 \bigg). \label{eq:14}
\end{eqnarray}
The field-theoretic description of dark energy coupled to dark matter treats both components as scalar fields. As the interaction is proposed between these fields, it is observed that the dark energy (DE) and dark matter (DM) do not satisfy the energy-momentum tensor conservation equation independently, instead, they satisfy the local conservation equation in the form given as:
\begin{equation}
\label{eqn:15}
    -\nabla^\mu T_{\mu \nu} ^{(\phi)} = Q_\nu= \nabla^\mu T_{\mu \nu} ^{(\psi)},
\end{equation}
where,
\begin{equation}
\label{eqn:16}
    Q_\nu= \nabla^\mu T_{\mu \nu} ^{(\psi)} = -\bigg(\frac{B,_\phi}{2} \nabla_\xi \psi \nabla^\xi \psi + 2 \alpha \phi \psi^2 \bigg) \nabla_\nu \phi,
\end{equation}
stands in for the energy transfer between dark energy and dark matter in interacting dark zone. Here, $T_{\mu \nu} ^{(\psi)}$ and $T_{\mu \nu} ^{(\phi)}$ play the role of energy-momentum tensor of field $\psi$ and field $\phi$, respectively.\\
Now, evolved from Eq.\eqref{eqn:16}, we can write the conservation equations for the fields $\phi$ and  $\psi$ in FLRW space-time, respectively, as:
\begin{eqnarray}
-\ddot{\phi}-3H\dot{\phi}-\frac{B,_\phi}{2} \dot{\psi}^2 + 2 \alpha \phi \psi^2 + V,_\phi = 0,  \label{eq:17} \\
\ddot{\psi}+3H\dot{\psi}+\frac{B,_\phi}{B} \dot{\phi} \dot{\psi} + 2 \alpha \psi \phi^2 B^{-2} = 0 . \label{eq:18}
\end{eqnarray}

\section{Similitude between field theory strategy and phenomenological fluid strategy  }  \label{4}
This section outlines one-to-one scaling between the field theory approach and the fluid approach of the interacting dark sector. This type of work has also been modeled and studied before in \cite{PhysRevD.103.023510}. We, here, try to illustrate that the interaction function $Q_\nu$, achieved with the assistance of classical field theory exhibits a unique form when the fluid approach has been applied to the interacting dark sector model under consideration. Also, the above-mentioned interaction function $Q_\nu$ has been derived systematically from the series of equations starting from action integral to conservation equations where one of the fields becomes dark matter representative.
\par
Now, starting up with the fluid description of the above interacting DE-DM model, one of the fields can be described as phenomenological fluid. In such description, it is often suitable to consider the dark matter as a fluid component. The openness in the work provides the freedom where we can assume either the field ``$\phi$" or the field ``$\psi$" as a dark matter entity. Therefore, we substitute the dark matter fluid for the scalar field ``$\psi$". Thus, framing the field ``$\phi$" as a DE entity. In this description, the dark matter energy density $\rho_{dm}$ and pressure $P_{dm}$ are given as follows:
\begin{eqnarray}
\rho_{dm}=-\frac{1}{2} B(\phi) \bigg(g^{\mu \nu} 
\nabla_{\mu} \psi \nabla_{\nu} \psi - 2 \alpha \phi^2 \psi^2 B^{-1} \bigg), \label{eq:19}\\
P_{dm}=-\frac{1}{2} B(\phi) \bigg(g^{\mu \nu} 
\nabla_{\mu} \psi \nabla_{\nu} \psi + 2 \alpha \phi^2 \psi^2 B^{-1} \bigg). \label{eq:20}
\end{eqnarray}
whereas the energy-momentum tensor \eqref{eqn:6} for the DDI model in terms of DE scalar field ($\phi$) and DM fluid can be revised as:
\begin{equation}
\label{eqn:21}
    T_{\mu \nu}= \bigg(-\nabla_{\mu} \phi \nabla_{\nu} \phi + \frac{1}{2} g_{\mu \nu} \nabla_\xi \phi \nabla^\xi \phi - g_{\mu \nu} V(\phi) + P_{dm}  g_{\mu \nu} + ( \rho_{dm} + P_{dm})  u_{\mu}  u_{\nu}\bigg),
\end{equation}
where $u_{\mu}$ is termed as the four-velocity of the DM fluid.\\
As observed in the literature, the phenomenological fluid models state the interaction term ``$Q_{\nu}$'' as a linear function of DM density or DE density or both. Also, it has been seen that the ``$Q_{\nu}$'' in fluid models is not uniquely acquired considering that it is simply set by hand. Unlike in fluid models, the interaction term ``$Q_{\nu}$'' derived in \eqref{eqn:22} has a particular form that is acquired by correlating the fields and fluids for the present interacting model.
\par
This analogy between fields and fluids assists us to revised all the dynamical field equations and conservation equations in terms of dark matter fluid component and dark energy field component for the DDI model. Here, in our work, we consider dark matter as a pressure-less entity.
\\
Now, from Eq. \eqref{eqn:16}, interaction function $ Q_\nu$ can be revised as:
\begin{equation}
\label{eqn:22}
    Q_\nu = \bigg( \frac{B,_{\phi}}{B}  -  \frac{2}{\phi} \bigg)\frac{\rho_{dm}}{2} \nabla_{\nu} \phi,
\end{equation}
and the Friedmann equations and the conservation equations in the FLRW space-time then revised, respectively, as in \eqref{eqn:23} and \eqref{eqn:24}
\begin{equation}
\label{eqn:23}
\begin{split}
3H^2= \rho_{dm} - \frac{\dot{\phi}^2}{2}+ V(\phi), \\[4pt]
3H^2 + 2\dot{H}=-\bigg(P_{dm}-
\frac{\dot{\phi}^2}{2}- V(\phi) \bigg).
\end{split}
\end{equation}
\begin{equation}
\label{eqn:24}
\begin{split}
-\ddot{\phi} \dot{\phi}-3H\dot{\phi}^2+V,_\phi \dot{\phi}= Q ,\\[4pt]
\dot{\rho}_{dm}+3H\rho_{dm}=-Q .
\end{split}
\end{equation}
Here, it is noted that the DE component $``\phi"$ has negative kinetic term, thus it is phantom field. Though the phantom field $``\phi"$ exhibits instabilities in quantum field arena \cite{CALDWELL200223,Hao:2003ib}, could be stable in classical field theory and may clearly be represented by a scalar field with negative kinetic energy term (Eq.\ref{eqn:23}). In such models, the evolution of the universe is phantom dominated with $\omega_\phi < -1$ \cite{PhysRevD.68.023522}. It is interesting to know that the cosmological evolution of $\omega_\phi$ (w.r.t. redshift(z)) less than -1 is very much acknowledged by the observations \cite{Hannestad:2002ur,Melchiorri:2002ux,Lima:2003dd,Alam:2004jy,Alam:2003fg,Wang:2003gz}. Though this feature can exhibit an unusual dynamics as it violates the energy conditions \cite{CALDWELL200223,Santos:2006ja,Carroll:2003st,Visser:1999de}, the exotic form of dark energy in terms of phantom fields are profoundly studied in the literature seeing the approval from latest observational findings \cite{Alam:2003fg,Wang:2003gz,Carroll:2003st} and leads to the required current accelerated expansion of the universe. (Nevertheless, it has been investigated \cite{BARROW1988743,CALDWELL200223} that it is not harmless for an energy component to violate the dominant energy condition (DEC) or weak energy condition (WEC) that too for a finite time period.) So, here, the motivation to consider such an exotic form of energy is strongly driven by the observations. 

It can also be realized from the EoS parameter less than -1 that the phantom energy density may rise to diverge with time, possibly causing  big rip singularity in finite time in future \cite{PhysRevD.71.063004,Sami:2003xv}. However, it has been shown by various studies that this can be over-passed, for details see \cite{ASTASHENOK2012396,Nojiri:2009pf}.
However, in this work, we have shown that the problem of big rip singularity does not occur. \\ 
\section{Dynamical system formulation}  \label{5}
The cosmological equations pertaining to the evolution of an isotropic and homogeneous universe are purely a network of ordinary differential equations. An impressive technique to study such networks is by moulding them into an autonomous (dynamical) system. The mathematical background behind the dynamical systems helps us to realize the qualitative features of the cosmological model under analysis. To study the dynamics of the system, we present the following set of expansion normalized variables
\begin{align}
\label{eqn:25}
    x &=\frac{\dot{\phi}}{\sqrt{6}H},&
    y &=\frac{{\sqrt{V(\phi)}}}{\sqrt{3}H},&
    \Omega_{m} &=\frac{\rho_{dm}}{3H^2}.&
\end{align}
where, using Eq. \eqref{eqn:23} and Eq. \eqref{eqn:24} the equivalent form of the field equations for an interacting DE-DM model become:\\
\begin{eqnarray}
x^\prime= - \sqrt{\frac{3}{2}}\bigg(\lambda y^2 +\frac{I}{x}\bigg)-\frac{3}{2}x (1+x^2+y^2),  \label{eq:26}\\[0.5pt]
y^\prime= -\sqrt{\frac{3}{2}}(\lambda x y) +\frac{3}{2}y (1-x^2-y^2), \label{eq:27}\\[0.5pt]
\Omega_m '= -3 \Omega_m (x^2 + y^2) -\sqrt{6}I , \label{eq:28}\\[0.5pt]
\lambda' = \sqrt{6}  \lambda^2 x(1- \Gamma_\lambda), \label{eq:29}\\[0.5pt]
k' = \sqrt{6} k^2 x (1- \Gamma_k),  \label{eq:30}\\[0.5pt]
\beta' = \sqrt{6} \beta^2 x (1- \Gamma_\beta),  \label{eq:31}\\[0.5pt]
\alpha'= - \sqrt{6} \alpha \beta x. \label{eq:32}
\end{eqnarray}
in which $\lambda$, k, $\beta$ and $\alpha$ are defined as
%
%\begin{align}
%\label{eqn:33}
%    \lambda(\phi) &=-\frac{V,_\phi}{V(\phi)},&
%    k(\lambda) &=-\frac{B,_\phi}{B(\phi)},&
%    \beta(\phi) &=-\frac{\alpha,_\phi}{\alpha(\phi)},&
%    \alpha = \alpha(\phi).&
%\end{align}
%

%
\begin{equation}
\label{eqn:33}
        \lambda(\phi) =-\frac{V,_\phi}{V(\phi)}, \qquad k(\lambda) =-\frac{B,_\phi}{B(\phi)}, \qquad \beta(\phi) =-\frac{\alpha,_\phi}{\alpha(\phi)}, \qquad \alpha = \alpha(\phi).
\end{equation}
and the functions $\Gamma_\lambda,\Gamma_k,\Gamma_\beta$.  are defined as
\begin{equation}
\label{eqn:34}
        \Gamma_\lambda =\frac{{V,_{\phi \phi} V}}{V,_\phi^2},\qquad \Gamma_k =\frac{{B,_{\phi \phi} B}}{B,_\phi^2}, \qquad \Gamma_\beta=\frac{{\alpha,_{\phi \phi} \alpha}}{\alpha,_\phi^2}.
\end{equation}
Here, prime denotes derivative with respect to the number of e-foldings N = ln(a) and \hspace{0.1cm}`$,_{\phi}$' denotes the partial derivative w.r.t $\phi$.\\ We write the scaled interaction term (I) as
\begin{equation}
\label{eqn:35}
    \text{I}=\frac{(2 \beta - k)}{2}\Omega_m x \hspace{0.1cm} = \frac{1}{3}\frac{Q}{\sqrt{6}H^3}.
\end{equation}
where `$\Omega_m$' symbolizes the relative density parameter for the dark matter component. It has been shown in figure \ref{fig:2} that  $\Omega_m$ varies almost steadily at late times. Please note that this feature has been taken into account for the further phase-space analysis of the system in section \ref{5.3}.\\
Also, Eq. \eqref{eqn:25} provides us the Friedmann constraint in terms of density parameter for dark matter as
\begin{equation}
\label{eqn:36}
     1+x^2-y^2=\Omega_m.
\end{equation}
Now, we compile other cosmological parameters in terms of dynamical variables \eqref{eqn:25} .\\
The density parameter for the phantom field (DE) can be written as:
\begin{equation}
\label{eqn:37}
    \Omega_\phi \hspace{0.1cm}= -x^2+y^2,
\end{equation}
with the equation of state parameter $\omega_{\phi}$ as:
\begin{equation}
\label{eqn:38}
    \omega_\phi \hspace{0.1cm}= \frac{-x^2-y^2}{-x^2 +y^2}.
\end{equation}
Then the effective equation of state parameter for the DE-DM interaction model becomes:
\begin{eqnarray}
 \omega_{eff} \hspace{0.1cm}=\frac{P_{eff}}{\rho_{eff}} \hspace{0.1cm}= \frac{P_\phi}{\rho_\phi + \rho_m},  \label{eq:39}\\
\omega_{eff} \hspace{0.1cm}= -x^2 -y^2 , \label{eq:40}
\end{eqnarray}
and the deceleration parameter is:
\begin{equation}
\label{eqn:41}
    q \hspace{0.1cm}= -1+\frac{3}{2}(1+\omega_{eff}).
\end{equation}
It can be seen that the acceleration condition is feasible for negative ``q" values i.e., for q$<0$,  whereas the deceleration condition is feasible for positive ``q" values i.e., for q$>0$.

\subsection{Critical points and stability analysis}  \label{5.1}
Now, we submit the comprehensive phase-space study of the dynamical system \eqref{eq:26}-\eqref{eq:32} )(for such phase-space analysis, refer \cite{article, BAHAMONDE20181,PhysRevD.103.123517}). The 7D autonomous system for an interacting DE-DM model is very much tangled and analysing the stability of the critical points is again a trickier work. In order to facilitate the further analysis, an exponential potential of the form $V(\phi)=V_0 e^{\alpha (\phi)}=V_0 e^{c' \phi}$ is the most suitable choice \cite{article, BAHAMONDE20181}. However, for the current study, `$ \alpha (\phi)$' is taken to be a non-zero constant to additionally reduce the dimensionality of the system. By default Eq. \eqref{eq:29}, \eqref{eq:31} and \eqref{eq:32}, then becomes trivial fetching us an apt 4D autonomous system.\\ 
Though such a model has been considered before in \cite{PhysRevD.103.023510}, the rigorous fixed-point analysis and stability analysis for this distinct exemplar has not been performed. Such analysis is essential to understand the mystifying dynamics of the observable universe in depth. Following this, we now find out the critical points and discuss their existence and stability properties. The system contains four critical points `$A$', `$B$', `$C$' and `$D$' in the phase space depending on the values of `$\lambda$' and `$\beta$' as presented in the Table \ref{tab:table1}. In the following, we examine the existence and acceleration features of each of the fixed points corresponding to their eigenvalues in the parameter space ($\beta$, $\lambda$) as presented in Table \ref{tab:table2}.
\begingroup
\begin{table}[!ht]
\caption {\label{tab:table1} The critical points and cosmological parameters corresponding to an autonomous system of  an interacting DE-DM model with an exponential potential and constant $\alpha$.} 
\begin{center}
\begin{tabular}{c| c c c c c c}
\toprule
\\
 Critical\\ Points & x & y & $\Omega_m$ & $\Omega_\phi$ & k & $\omega_{eff}$ \\
\\ \bottomrule \\ 
 A & 0 & 0 & $\Omega_m$ & 0 &$2 \beta$ & 0        \\[0.4cm]     
 B & $-\sqrt{\frac{2}{3}} \beta$ & 0 & $1+\frac{2}{3} \beta^2$ & $-\frac{2}{3} \beta^2$ & 0  & $-\frac{2}{3} \beta^2$       \\[0.4cm]
 C & $\frac{-\sqrt{\frac{3}{2}}}{ \beta- \lambda}$ &
 $\frac{\sqrt{-\frac{3}{2}+ \beta^2 -  \beta \lambda}}{( \beta - \lambda)}$ & $\frac{3-  \beta \lambda + \lambda^2}{( \beta - \lambda)^2}$ & $\frac{-3+ \beta^2 -  \beta \lambda}{( \beta - \lambda)^2}$ &0 &  ${\frac{1}{\frac{\lambda}{ \beta}-1}}$\\[0.4cm]
 D & $\frac{-\lambda}{\sqrt{6}}$ & $\sqrt{1+\frac{\lambda^2}{6}}$ & 0 & 1 & 0 &$-1-\frac{\lambda^2}{3}$\\
\end{tabular}
\end{center}
\end{table}
\endgroup
\begin{itemize}
     \item \textbf{Critical point A} exist for all real values of parameter $\beta$. The point is completely scarce of scalar field $\phi$ as there is no contribution either from kinetic energy or from the potential energy of the field. For pressure-less DM, the effective EoS ($\omega_{eff}$) vanishes which implies that the point does not facilitate the condition for accelerating universe. Following the constraint equation \eqref{eqn:36}, the relative density parameter for DM becomes one and hence the point can be a representative of the DM-dominated universe in the past. The critical point A is of non-hyperbolic nature. For the condition, ($\beta$, $\Omega_m$) $\neq 0$, only one of the eigenvalues is zero [Table \ref{tab:table2}]. Thus, the critical line corresponding to the point A fits with the $\Omega_{m}$ axis and the imposition of the centre manifold theory becomes an impracticable job (For details, refer \cite{BAHAMONDE20181}, Sec. 4.4). This implies that for $\Gamma_k <1$, the critical line of the point A acts as a saddle line but for $\Gamma_k >1$, the eigenvalues show complex nature with the real parts predicting the stability of the point A as saddle one.
\end{itemize}
\begingroup
\begin{table}[!ht]
\caption {\label{tab:table2} The existence and eigenvalues of corresponding critical points.} 
\begin{center}
\begin{tabular}{c| c c c c c c}
\toprule \\
 Critical Pts. &  Existence &  Accel. & Eigenvalues  \\ 
 \\ \bottomrule \\
 A & $\forall  \beta$  & No & ${\frac{3}{2}, 0,-\frac{3}{4}\pm \frac{\sqrt{3}}{4}\sqrt{3+32 \beta^2 \Omega_m (1-\Gamma_k)}}$  \\ [0.4cm] 
 B & \begin{tabular}{@{}c@{}}$\forall \beta$ \end{tabular} 
  & Yes & $0, -2 \beta^2,-\frac{3}{2}-\beta^2,\frac{3}{2}-\beta^2+ \lambda \beta$ \\[0.4cm]
 C & $\frac{\lambda}{ \beta}> -2$ & Yes & See Appendix  \\[0.4cm]
 D & $\forall \lambda$ & Yes & ${0,-3-\frac{\lambda^2}{2}, -3-\lambda^2+ \beta \lambda,-3-\lambda^2, }$\\
 \end{tabular}
\end{center}
\end{table}
\endgroup
\begin{itemize}
     \item \textbf{Critical point B} is dominated by the kinetic energy of the scalar field $\phi$. Its existence, like point A, is valid for any real value of parameter $\beta$. The point pictures the existence of an accelerating solution ($\omega_{eff}<-\frac{1}{3}$), locating in the region of parameter space where the condition $ \beta^2 > \frac{1}{2}$ is being satisfied. This condition can lead us to both, quintessence and phantom dominated DE accelerating universe. The critical point B is also non-hyperbolic in nature like point A and the Centre Manifold Theory can be employed to reveal the nature of point B.
\end{itemize}

\begin{figure}[H]
  \centering
  \begin{subfigure}[b]{0.45\linewidth}
    \includegraphics[width=\linewidth]{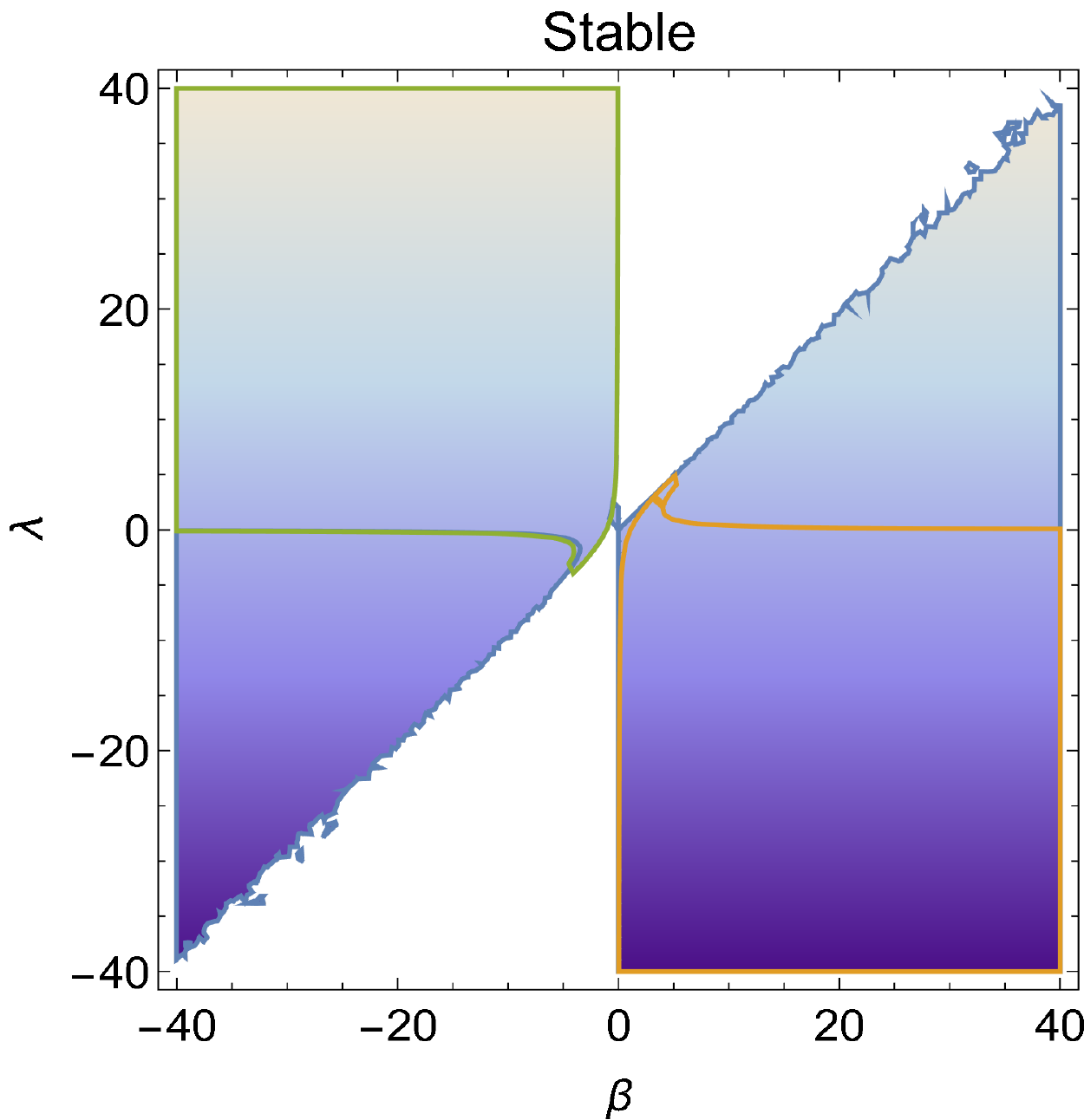}
     \caption{}
     \label{fig:region plot_stable}
  \end{subfigure}
  \begin{subfigure}[b]{0.45\linewidth}
    \includegraphics[width=\linewidth]{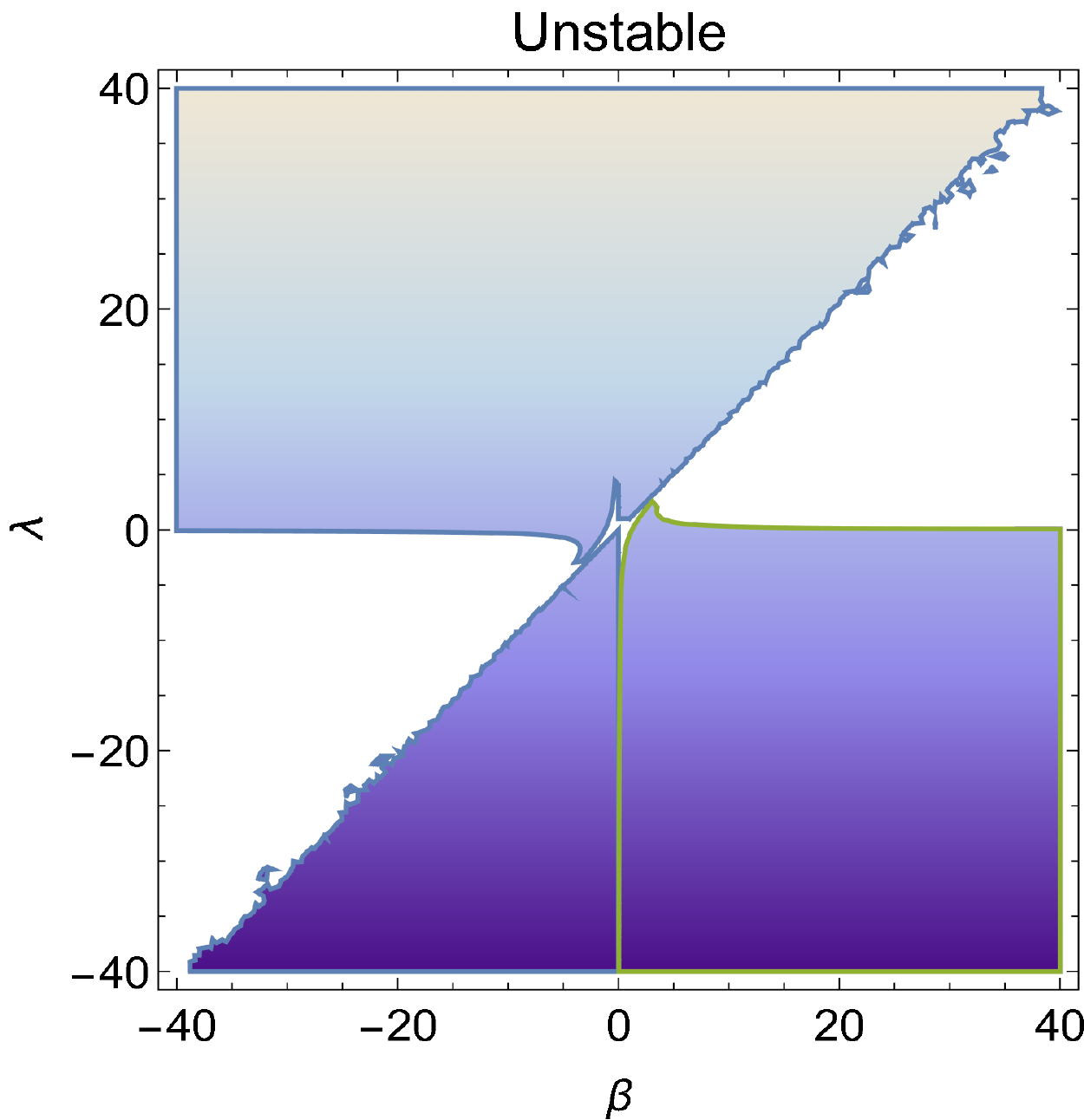}
    \caption{}
    \label{fig:region plot_unstable}
  \end{subfigure}
  \caption{The allowed range of parameter ${[\beta,\lambda]}$ space in which critical point C exhibits (a) stability and (b) instability characteristics.}
  \label{fig:region plot}
\end{figure}

\begin{itemize} 
     \item \textbf{Critical point C} is valid for $-2 < \frac{\lambda}{\beta}<1$ in the phase-space. It exhibits the scaling solution where the ratio between DE and DM density parameter throws a constant
     \begin{equation*}
         \frac{\Omega_{\phi}}{\Omega_m} = \frac{-3+ \beta^2-  \beta \lambda}{3+\lambda^2-  \beta \lambda}.
     \end{equation*}\\[4pt]
     \vspace{0.1cm}
     The accelerated evolution of the universe for the point C is possible within the above set limit for $\frac{\lambda}{ \beta}$. \\
     When $\frac{\lambda}{ \beta}$ becomes zero. The DE i.e. scalar field `$\phi$' portrays the cosmological constant behaviour ($\omega_{eff} = -1$) and an accelerated de Sitter universe is achieved.\\
     For $0< \frac{\lambda}{ \beta}<1$ case, $\omega_{eff}$ ranges as $\omega_{eff}<-1$, clearly showcasing the phantom DE dominated behaviour. Whereas, for $-2< \frac{\lambda}{ \beta}<0$ case, $\omega_{eff}$ ranges as $-1<\omega_{eff}< -\frac{1}{3}$ and suits the condition under which scalar field (DE) behaves as a quintessence field.\\
     Point C is of non-hyperbolic type. Stating this, it is found that the complexity of the eigenvalues (due to higher power parameters) and the eigenvectors of the point C obstructs us from carrying out further analysis, required to implement the Centre Manifold Theory to decide the nature of the fixed point. Thus, with a comprehensive purpose, to show the stable properties of this point we follow an analytical approach. We plot the parameter space regions for the fixed point to be stable and unstable in figure \ref{fig:region plot}. The figure \ref{fig:region plot_stable} portrays the regions of the parameter space where the point C is stable whereas the figure \ref{fig:region plot_unstable} portrays the regions of the parameter space where the point C exhibits instability and the intersection hyper-surface between these two regions represents the saddle surface.\\
     We now carry out further analysis for fixed-point C with the help of the table \ref{tab:table1}. As can be seen from the figure \ref{fig:region plot} that at some specific regions of parameter space where $`\lambda$' vanishes and $`\beta$' is non-negative, the kinetic energy becomes negative and the universe exhibits phantom dominated behaviour. On the contrary, in the regions where $`\beta$' becomes zero and $`\lambda$' is non-negative, the kinetic term becomes canonical. Thus, in order to be dominated by the standard canonical kinetic term, the point needs to be unstable. Hence, in the ongoing case, we conclude that though the flow becomes dynamic and highly unstable, slight fluctuation will resurrect the stability. The above conclusion substantiates the choice of action surveyed in the present study, where the fusion of canonical and non-canonical scalar fields, with little fluctuation, can introduce both stability and instability dynamics for the zeroth-order background universe.
\end{itemize}
\begin{itemize}
     \item \textbf{Critical point D} is completely dominated by scalar field $\phi$ $(\Omega_{\phi} =1)$ in the phase-space. The point D exists for all real values of $\lambda$. For $\lambda^2 >-2$ limit, it describes an accelerating solution.\\
     The case in which $\lambda=0$, the scalar field (DE) behaves as cosmological constant ($\omega_{eff} = -1$). For the case $\lambda^2>0$, the universe portrays phantom field dominated behavior. Whereas the case $-2<\lambda^2 <0$, portrays the quintessence field dominated behavior.\\
     Point D is also a non-hyperbolic type critical point. As explained in the above case for point B, the linear stability theory is not sufficient to determine its nature. Therefore, we apply the centre manifold theory to study the nature of point D.
\end{itemize}
 \subsection{Centre manifold theory}  \label{5.2}
We now examine the stability properties of the above fixed points B and D.
\begin{center}
    \textbf{Critical point B}
\end{center} 
The Jacobian matrix at the point B can be set as
\begin{equation}
\label{eqn:42}
J_B=
\begin{bmatrix}
-\frac{3}{2}-3\beta^2 & 0 & -\sqrt{\frac{3}{2}} \beta & \sqrt{\frac{3}{2}}\left(\frac{1}{2}+\frac{\beta^2}{3} \right) \\[0.2cm]
0 & \frac{3}{2}-\beta^2+\beta \lambda & 0 & 0\\[0.2cm]
\sqrt{\frac{2}{3}} \beta(3+2 \beta^2) & 0 & 0 & -\beta\left(1+\frac{2}{3}\beta^2\right)\\[0.2cm]
0 & 0 & 0 & 0
\end{bmatrix}
\end{equation}
\vspace{0.3cm}
So the eigenvalues of J$_B$ are given as $0,\hspace{0.1cm}-2\beta^2,\hspace{0.2cm}-\frac{3}{2}-\beta^2,\hspace{0.2cm}\frac{3}{2}-\beta^2+\beta \lambda$ \\
\vspace{0.2cm}
and the corresponding eigen vectors are
$\left[\frac{1}{\sqrt{6}}, \hspace{0.1cm}0, \hspace{0.1cm} \frac{-2\beta}{3}, \hspace{0.1cm} 1\right]^T$,
$\left[-\frac{\sqrt{6}\beta}{3+2\beta^2},\hspace{0.1cm}0,\hspace{0.1cm}1,\hspace{0.1cm}0\right]^T$, 
$\left[-\frac{\sqrt{\frac{3}{2}}}{2\beta},\hspace{0.1cm}0,\hspace{0.1cm}1,\hspace{0.1cm} 0\right]^T$ and \hspace{0.1cm}[0,\hspace{0.1cm}1,\hspace{0.1cm}0,\hspace{0.1cm}0]$^T$.

\vspace{0.3cm}
To apply centre manifold theory, we first perform coordinate transformations such that the critical point moves to the origin and the system, accordingly changing to
\begin{equation*}
    x =X-\sqrt{\frac{2}{3}} \beta, \qquad y=Y, \qquad z=Z+\left(1+\frac{2}{3} \beta^2\right), \qquad k=K.
\end{equation*}
Here, the parameter $\Omega_m$ is replaced with $`z$' to ease the symbolic notation.\\
We bring in one more set of coordinates ($x_t,y_t,z_t,k_t$) in terms of ($X,Y,Z,K$). By using the eigenvectors of the J$_B$, we insert the following new coordinate system
\begin{center}
\begin{equation}
\label{eqn:43}
\begin{bmatrix}
x_t\\[0.2cm]  y_t\\[0.2cm]  z_t\\[0.2cm]  k_t
\end{bmatrix}=
\begin{bmatrix}
\frac{\sqrt{6}\beta }{3+2\beta^2} & 0 & \frac{\sqrt{\frac{3}{2}}}{2\beta} & -\frac{1}{\sqrt{6}} \\[0.2cm]
0 & 1 & 0 & 0\\[0.2cm]
1 & 0 & 1 & \frac{2\beta}{3}\\[0.2cm]
0 & 0 & 0 & 1
\end{bmatrix}
\begin{bmatrix}
X\\[0.2cm]  Y\\[0.2cm]  Z\\[0.2cm]  K
\end{bmatrix}
\end{equation}
\end{center}
\vspace{0.5cm}
This new coordinate system allows us to transform the autonomous system \eqref{eq:26}-\eqref{eq:28} and \eqref{eq:30} as
\begin{center}
\begin{equation}
\label{eqn:44}
\begin{bmatrix}
x'_t\\[0.2cm]  y'_t\\[0.2cm]  z'_t\\[0.2cm]  k'_t
\end{bmatrix}=
\begin{bmatrix}
A(\beta) & 0 & 0 & 0 \\[0.2cm]
0 & (\frac{3}{2}-\beta^2+\beta \lambda) & 0 & 0\\[0.2cm]
0 & 0 & B(\beta) & 0\\[0.2cm]
0 & 0 & 0 & 0
\end{bmatrix}
\begin{bmatrix}
x_t\\[0.2cm]  y_t\\[0.2cm]  z_t\\[0.2cm]  k_t
\end{bmatrix}+
\begin{bmatrix}
non\\[0.1cm]linear\\[0.1cm] terms
\end{bmatrix}
\end{equation}
\end{center}
where,
\begin{align*}
    A(\beta) &=  \frac{\beta}{(-9+6\beta^2)}\bigg[9\sqrt{6}+2\beta(-9+\beta(9\sqrt{6}+2\beta(-9+\sqrt{6}\beta))) \bigg]  ,\\
    B(\beta) &=\frac{1}{6(-3+2\beta^2)}\bigg[27-2\beta(9\sqrt{6}+2\beta(-18+\beta(9\sqrt{6}+\beta(-9+2\sqrt{6}\beta)))) \bigg] .\\
    \\
\end{align*}
To find a definitive solution to these equations using standard methods is an impassable task. Thence, we make a series expansion of h($k_t$) in the powers of $k_t$.
By the definition of the centre manifold (refer to sec.2.4 of \cite{BAHAMONDE20181}), there exists a sufficiently regular function `h' such that 

\begin{center}
\begin{equation}
\label{eqn:45}
\begin{bmatrix}
x_t\\[0.3cm]
y_t\\[0.3cm]
z_t\\  
\end{bmatrix}=
h(k_t)
=
\begin{bmatrix}
a_1 k{^2_t}+a_2 k{^3_t}+\mathcal{O}({k{_t}^4})\\[0.3cm]
b_1 k{^2_t}+b_2 k{^3_t}+\mathcal{O}({k{_t}^4})\\[0.3cm]
c_1 k{^2_t}+c_2 k{^3_t}+\mathcal{O}({k{_t}^4}) \\
\end{bmatrix}
\end{equation}
\end{center}
\vspace{0.3cm}
We differentiate this with respect to `N' and apply the chain rule which
yields
\begin{equation}
\label{eqn:46}
    \begin{bmatrix}
    \frac{d x_t}{d N} \\[0.5cm]
    \frac{d y_t}{d N}\\[0.5cm]
    \frac{d z_t}{d N}
    \end{bmatrix}=
    \begin{bmatrix}
    2a_1 k_t+3a_2 k{^2_t}+\mathcal{O}({k{_t}^3})\\[0.5cm]
    2b_1 k_t+3b_2 k{^2_t}+\mathcal{O}({k{_t}^3})\\[0.5cm]
    2c_1 k_t+3c_2 k{^2_t}+\mathcal{O}({k{_t}^3})
    \end{bmatrix}
    \times
    \begin{bmatrix}
    \frac{d k_t}{d N} 
    \end{bmatrix}
\end{equation}
where, $a_i, b_i, c_i \in \mathbb{R}$.
As we analyze arbitrary small neighbourhood of the origin, we keep only lowest power terms in CMT. Comparing lowest powers of non-zero coefficients of $k_t$ from both sides of Eq. \eqref{eqn:46}, we deduce the solution as
$a_1=C(\beta, \Gamma)$, $b_i=0$ and $c_1=D(\beta, \Gamma)$ where, $C(\beta, \Gamma)$ and $C(\beta, \Gamma)$ can be interpreted from Eq.\eqref{eq:47} and Eq.\eqref{eq:49} respectively.\\
Thus, we observe that the centre manifold given by the Eq. \eqref{eqn:45} can be put down as
\begin{align}
 x_t &= \frac{1}{6} \bigg(-6-8 \Gamma_k -\frac{3}{\beta^2} - \frac{\sqrt{6}(3+\Gamma_k)}{\beta} + \frac{-6(7+10 \Gamma_k)+4 \sqrt{6}(1+4\Gamma_k) \beta}{-3+2\beta^2} + \frac{48(6+\sqrt{6} \beta)}{(3-2\beta^2)^2} \nonumber \\[0.5cm]
    & + \frac{18(-1+2\Gamma_k)}{(3+2 \beta^2)^2} + \frac{2(6+\sqrt{6}\beta + 2\Gamma_k(-6+ \sqrt{6}\beta))}{(3+2 \beta^2)^2} \bigg)k{^2_t}
    +\mathcal{O}(k{_t}^3) , \label{eq:47} \\[0.5cm]
 y_t &= 0 , \label{eq:48}\\[0.5cm]
 z_t &= \frac{1}{9}\bigg(
    6(-5+9\Gamma_k)-\frac{3\sqrt{6}}{\beta}-2\sqrt{6}(7+4\Gamma_k)\beta-\frac{3(-1+2\Gamma_k)(6+\sqrt{6}\beta)}{(3+2\beta^2)}  \nonumber  \\[0.5cm]
    &  +\frac{6(30-24\Gamma_k+15\sqrt{6}\beta)+10\sqrt{6}\Gamma_k \beta)}{(3-2\beta^2)} -\frac{144(3+2\sqrt{6}\beta)}{(3-2\beta^2)^2}\bigg) k{^2_t} +\mathcal{O}({k{_t}^3}) . \label{eq:49} 
\end{align}
\vspace{0.1cm}
and therefore, the flow on the centre manifold obtains the form as
\begin{eqnarray}
\frac{d k_t}{d N} = -2(1-\Gamma_k)\beta k{^2_t} + \frac{4\sqrt{6}(1-\Gamma_k)\beta(3+2\beta^2)}{-9+6\beta^2}k{^3_t}+\mathcal{O}({k{_t}^4}) . \label{eq:50}
\end{eqnarray}
We are interested only in the non-zero coefficients of lowest power terms of $k_t$ in CMT as we analyze an arbitrarily small neighbourhood of the origin. Accordingly, the lowest power term of the expressions of the center manifold is $k{^2_t}$ which depends upon the sign of $\Gamma_k$ and $\beta$. For the defined range of $\phi :0\backsim \pi$, the value of $\Gamma_k$ does not exceed 1, and that keeps $(1-\Gamma_k)$ a positive quantity during the complete analysis of point B. By looking at the centre manifold equations, it has been noted that an analytical approach to understand the stability and qualitative nature in the neighbourhood of the non-hyperbolic point B will be unusable, and thus, we aim to achieve the same with numerical analysis. For this purpose, we choose the different values of $\Gamma_k$ and $\beta$ with the already prescribed range of $\Gamma_k$. The various choices of these parameter values direct us towards the same form of the centre manifold and hence, showing the stability analysis only for one such set of values would be enough to understand its behaviour on the center manifold near the origin.
For the chosen value of $\beta= \pm0.5$ and $\Gamma_k =-1$ the stability analysis follows as:\\
\begin{figure}[!ht]
  \centering
  \begin{subfigure}[b]{0.46\linewidth}
    \includegraphics[width=\linewidth]{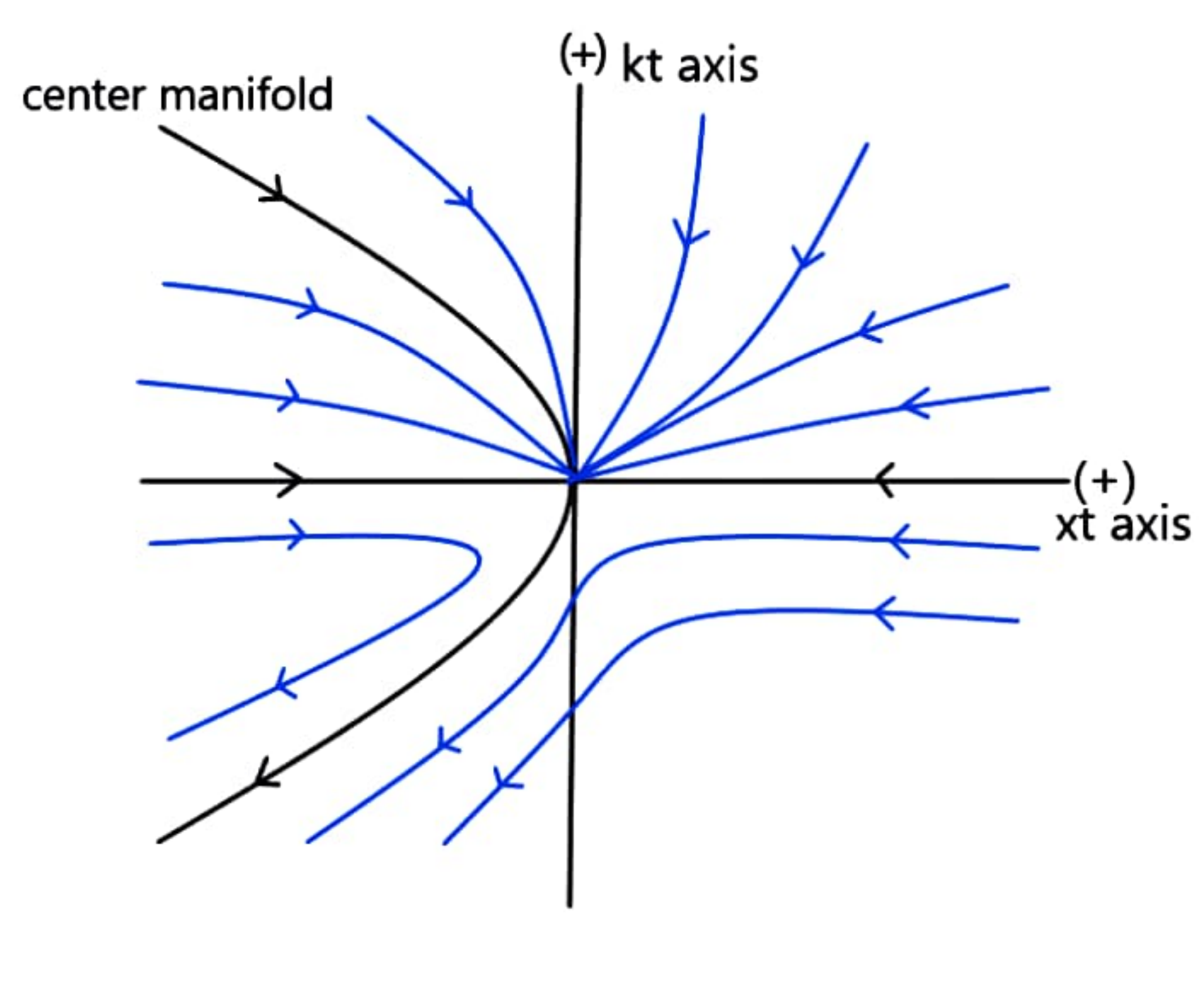}
     \caption{}
     \label{fig:vectorplot1}
  \end{subfigure}
  \begin{subfigure}[b]{0.46\linewidth}
    \includegraphics[width=\linewidth]{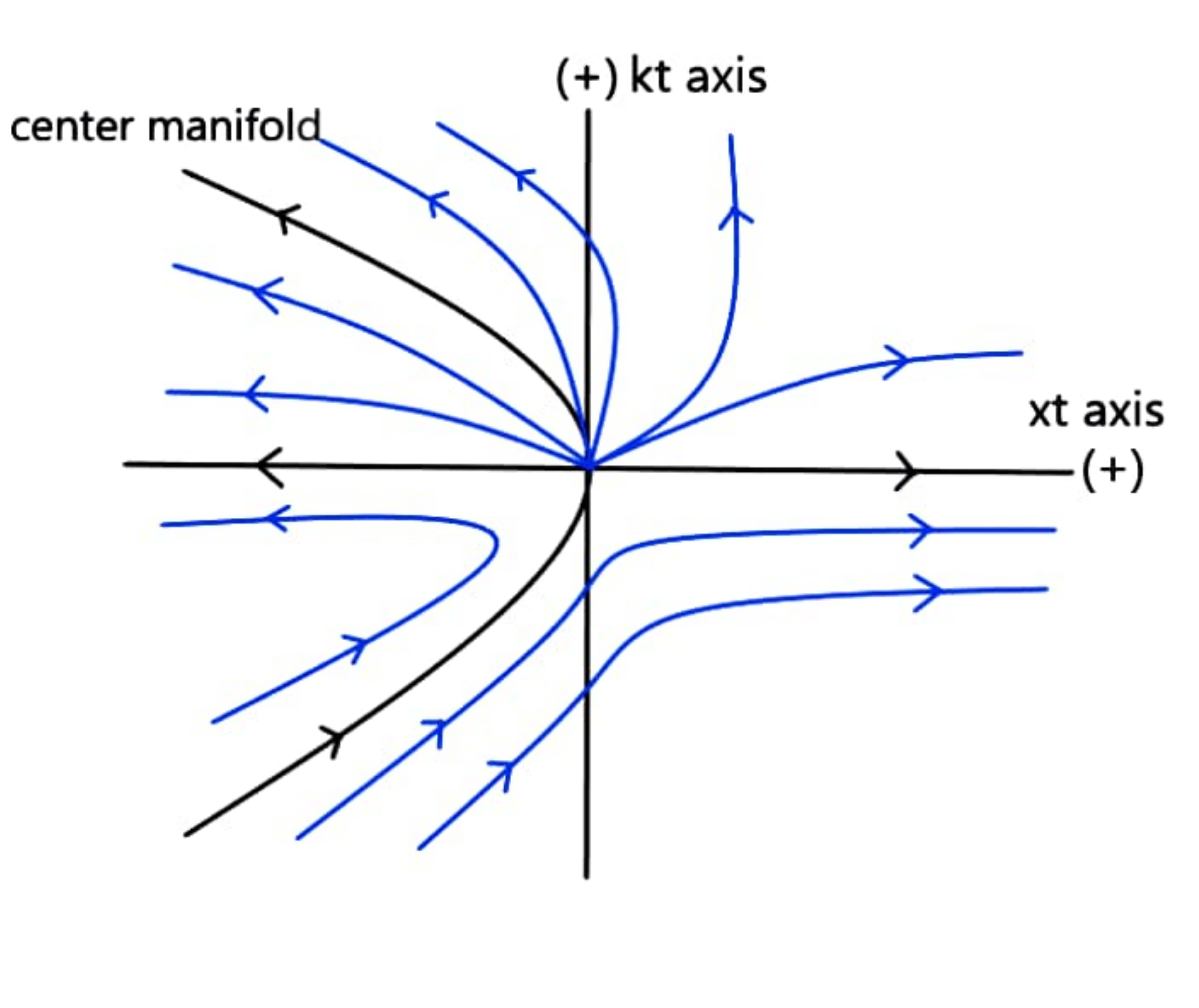}
    \caption{}
    \label{fig:vectorplot2}
  \end{subfigure}
  \begin{subfigure}[b]{0.46\linewidth}
    \includegraphics[width=\linewidth]{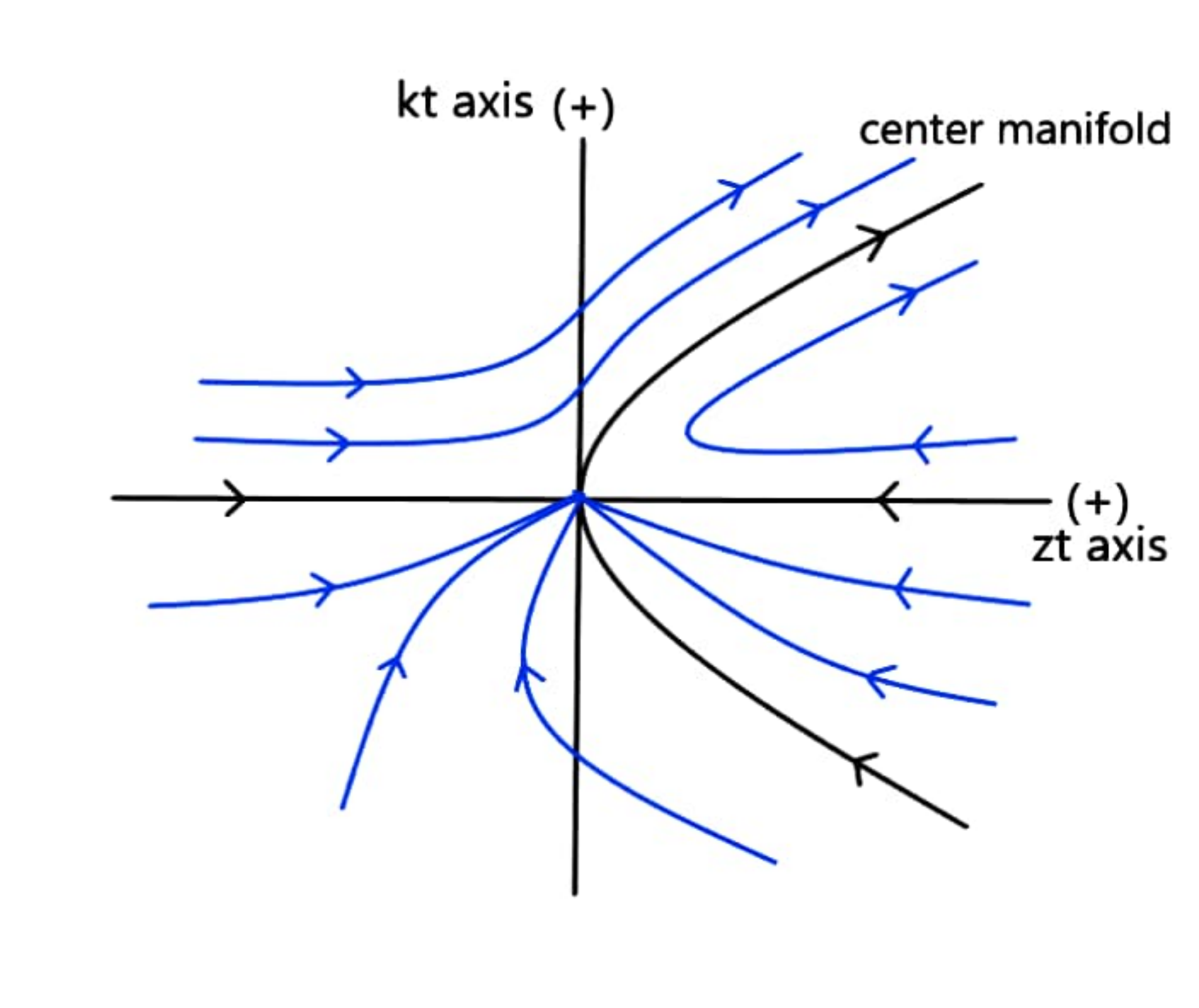}
    \caption{}
    \label{fig:vectorplot3}
  \end{subfigure}
  \caption{The phase portrait in the proximity of the origin for the critical point B. (a) represents the vector field near the origin in $x_t-k_t$ plane for $\beta > 0$. (b) and (c) respectively, represent the vector field near the origin in $x_t-k_t$ and $z_t-k_t$ plane for $\beta < 0$. The arrows indicate the flow along the center manifold (black curve). }
  \label{fig:criticalpointB_vectorfield}
\end{figure}
If $\beta>0$ then the origin becomes a saddle node and it is unstable. This signals that the local phase portrait is as given by figure \ref{fig:vectorplot1}. The behaviour of the vector field near the origin and the flow on the centre manifold in the $z_t-k_t$ plane is the same as figure \ref{fig:vectorplot1}. If $\beta<0$ then the local phase portrait is as given by figure \ref{fig:vectorplot2} and figure \ref{fig:vectorplot3} for $x_t-k_t$ and $z_t-k_t$ planes respectively. We detect that the origin is a saddle node and hence it is unstable in nature. The topological equivalence makes the behaviour of the non-hyperbolic point B in the old coordinate system similar to the behaviour of the centre manifold in the vicinity of the origin in the new coordinate system and that it is unstable due to its saddle nature.

\begin{center}
    \textbf{Critical point D}
\end{center} 
The Jacobian matrix at the point D can be set as
\begin{equation}
\label{eqn:51}
J_D=
\begin{bmatrix}
-3-\lambda^2 & -\frac{\lambda}{2} \sqrt{6+\lambda^2} & -\sqrt{\frac{3}{2}}  \beta & 0 \\[0.2cm]
0 & -3-\frac{\lambda^2}{2} & 0 & 0\\[0.2cm]
0 & 0 & (-3-\lambda^2+ \beta \lambda) & 0\\[0.2cm]
0 & 0 & 0 & 0
\end{bmatrix}
\end{equation}
\vspace{0.2cm}
So the eigenvalues of J$_D$ are given as $0,-3-\frac{\lambda^2}{2},-3-\lambda^2+\beta \lambda,-3-\lambda^2$ and \\
\vspace{0.1cm}
the corresponding eigen vectors are \hspace{0.1cm}
$\big[1,\hspace{0.1cm}0,\hspace{0.1cm}0,\hspace{0.1cm}0 \big]^T$, $\big[-\sqrt{1+\frac{6}{\lambda^2}},\hspace{0.1cm}1,\hspace{0.1cm}0,\hspace{0.1cm}0 \big]^T$, \hspace{0.1cm}
$\big[-\sqrt{\frac{3}{2\lambda^2}},\hspace{0.1cm}0,\hspace{0.1cm}1,\hspace{0.1cm}0 \big]^T$ and \hspace{0.1cm}
$\big[0,\hspace{0.1cm}0,\hspace{0.1cm}0,\hspace{0.1cm}1\big]^T$.\\
We, now modify the coordinates such that the critical point moves to the origin and the system, correspondingly changes to
\begin{equation*}
    x=X-\frac{\lambda}{\sqrt{6}}, \qquad y=Y+\sqrt{1+\frac{\lambda^2}{6}}, \qquad z =Z, \qquad k= K
\end{equation*}

Here as well, the parameter $\Omega_m$ is replaced with $`z$' to ease the symbolic notation.\\
Subsequently, settling the similar arguments as implemented in the above case, we understand that the centre manifold, in this case too, is given as
\begin{eqnarray}
 x_t = 0 , \label{eq:52}\\
 y_t = 0 , \label{eq:53}\\
 z_t = 0 . \label{eq:54}
\end{eqnarray}
and thus, making the flow on the centre manifold vanish as
\begin{equation}
\label{eqn:55}
 \frac{d k_t}{d N} = 0   .
\end{equation}
In this event, the data is not adequate to know the stability properties of point D near the origin. So, a bit different from the above analysis, we attempt to picture the stability of the vector field near the origin on each plane. The stability properties of the vector field on each plane for point D are shown in
Table \ref{tab:table4}. In the table, we denote, \hspace{0.1cm} $c=  \beta \lambda-\lambda^2$.
\vspace{0.2cm}
\begingroup
\begin{table}[h!]
\caption {\label{tab:table4} Stability properties of vector field for critical point D.}
\begin{center}
    \begin{tabular}{ | c |  c |}
      \hline
      \toprule
      \thead{Coordinate\\ plane} &  \thead{Stability}   \\
      \bottomrule
      \hline
      $x_t y_t$ &  \makecell{Vector field is stable about $x_t$ axis for any real value of $\lambda$.}    \\
      $x_t z_t$ &  \makecell{For $-3<-c$, vector field is stable about $x_t$ axis,\\ for $-3>-c$, vector field is unstable about $x_t$ axis,\\ for $c=3$, vector fields are nearly parallel to $x_t$ axis and obeys\\ stability behaviour about $k_t$ axis as shown in the figure \ref{fig:critical_point_D_vectorfield_x_t-z_t}.\\ In the plot the tracking nature of solutions running from higher $z_t$\\ to lower $z_t$ values is observed.}\\
      $x_t k_t$ &  \makecell{ Vector field is stable about $x_t$ axis for any value of $\lambda$.}\\
      $y_t z_t$ &  \makecell{For $-3<-c$, vector field near the origin is a stable node,\\ for $-3>-c$, vector field near the origin is a saddle node,\\ for $c=3$, vector field is stable about $z_t$ axis. }\\
      $y_t k_t$ & \makecell{Vector field near the origin acts as stable star for any real value of $\lambda$.}\\
      $z_t k_t$ & \makecell{For $-3<-c$, vector field near the origin is a stable node,\\ for $-3>-c$, vector field near the origin is a saddle node,\\ for $c=3$, vector field is stable about $z_t$ axis.}\\ \hline
    \end{tabular}
  \end{center}
  \end{table}
\endgroup
\begin{figure}[!ht]
    \centering
    \includegraphics[width=0.55\textwidth]{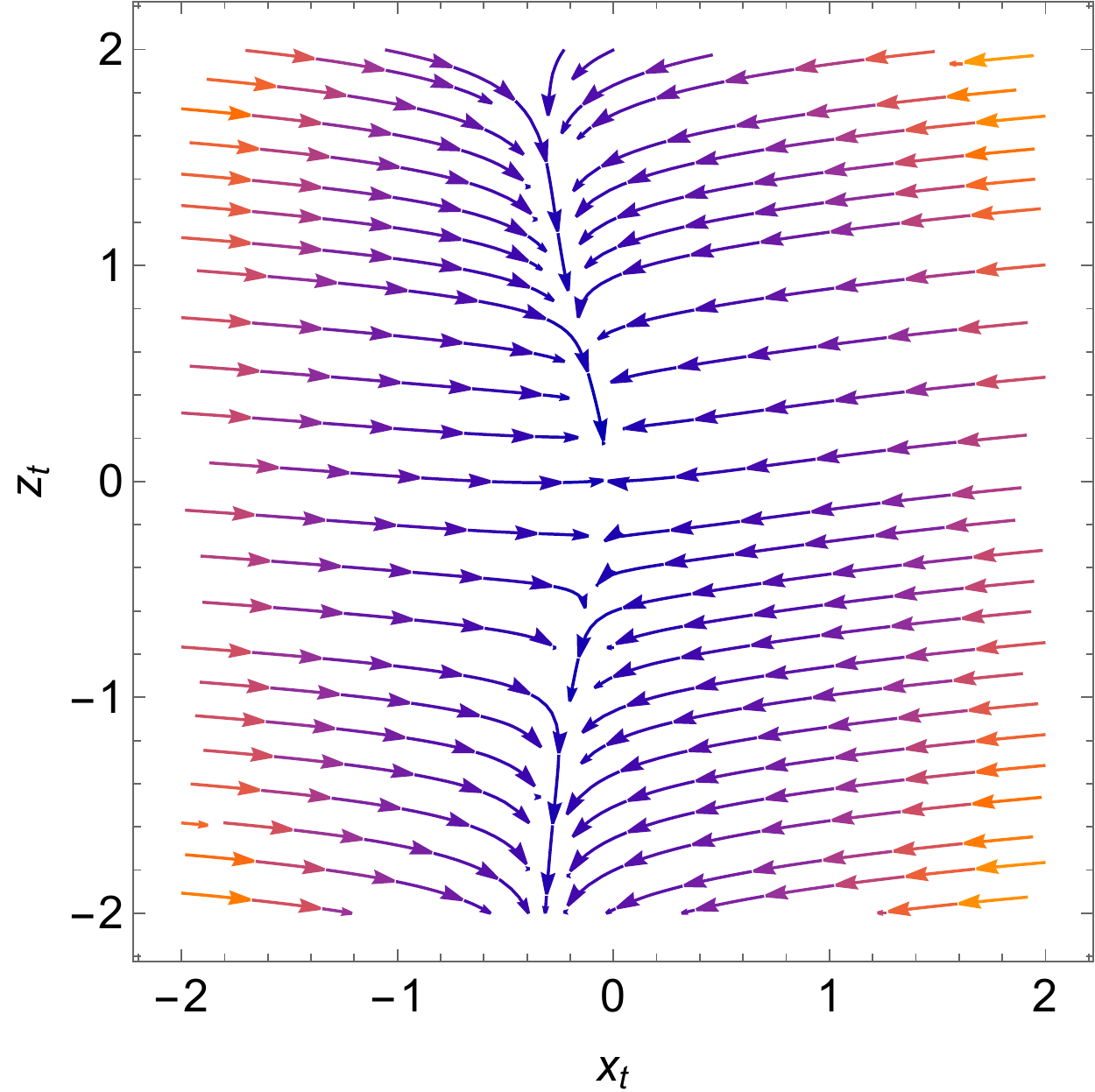}
    \caption{Vector field projection on $x_t z_t$ coordinate plane when $c=3$  for fixed point D for $\lambda=0.5$ and coupling constant `C'= 0.04. }
    \label{fig:critical_point_D_vectorfield_x_t-z_t}
\end{figure}\\
In the following subsection, we exhibit the graphical representation of the above qualitative analysis of the critical points with constraints on the DE-DM interaction model.

\subsection{Constraints on DE-DM interaction model}  \label{5.3}
\begin{figure}[!ht]
  \centering
  \begin{subfigure}[b]{0.49\linewidth}
    \includegraphics[width=\linewidth]{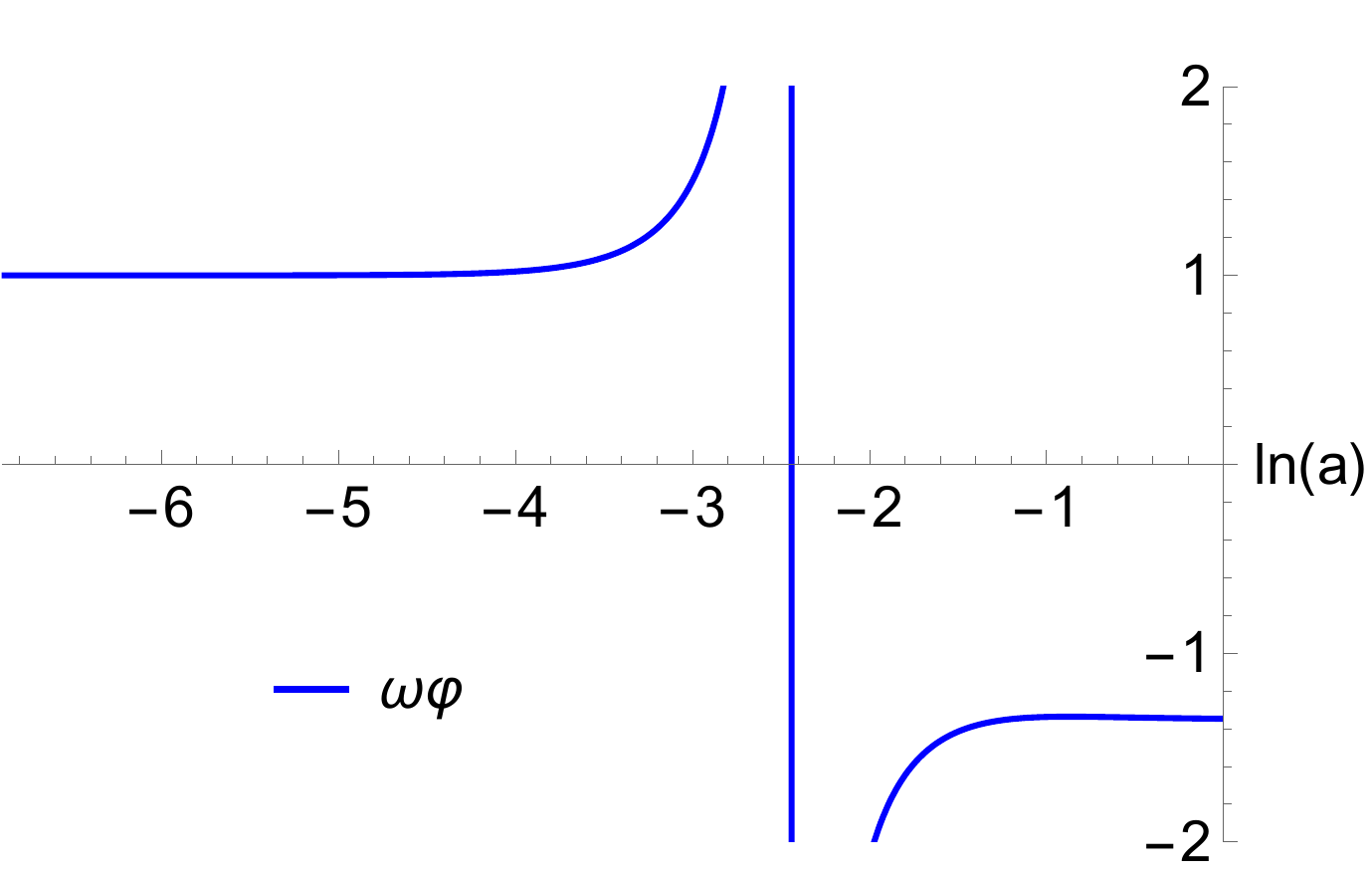}
     \caption{C = 0.8}
     \label{fig:wphi_c_0.8}     
  \end{subfigure}
  \begin{subfigure}[b]{0.47\linewidth}
    \includegraphics[width=\linewidth]{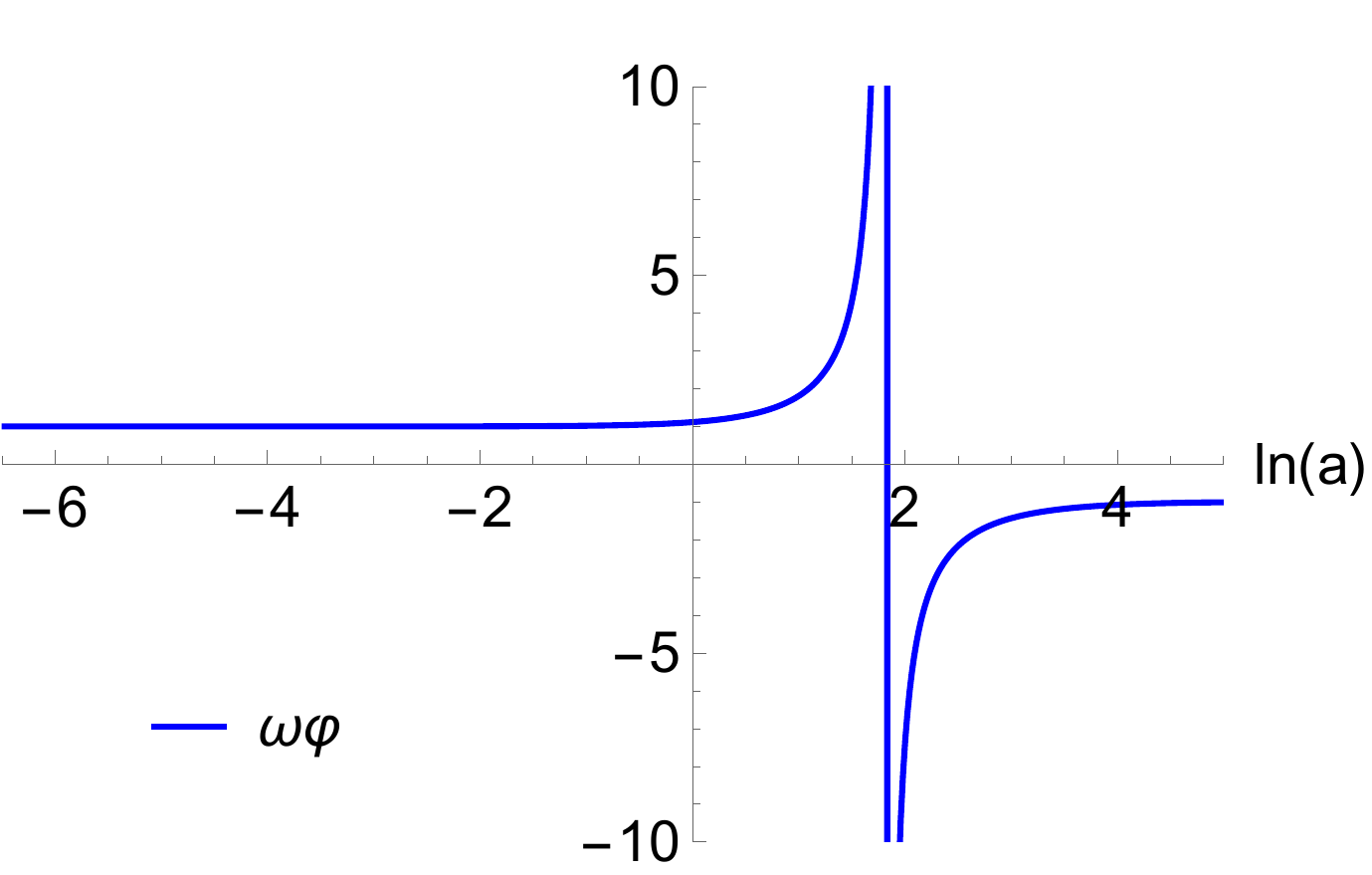}
    \caption{C = 1.2}
    \label{fig:wphi_c_1.2}
  \end{subfigure}
  \begin{subfigure}[b]{0.47\linewidth}
    \includegraphics[width=\linewidth]{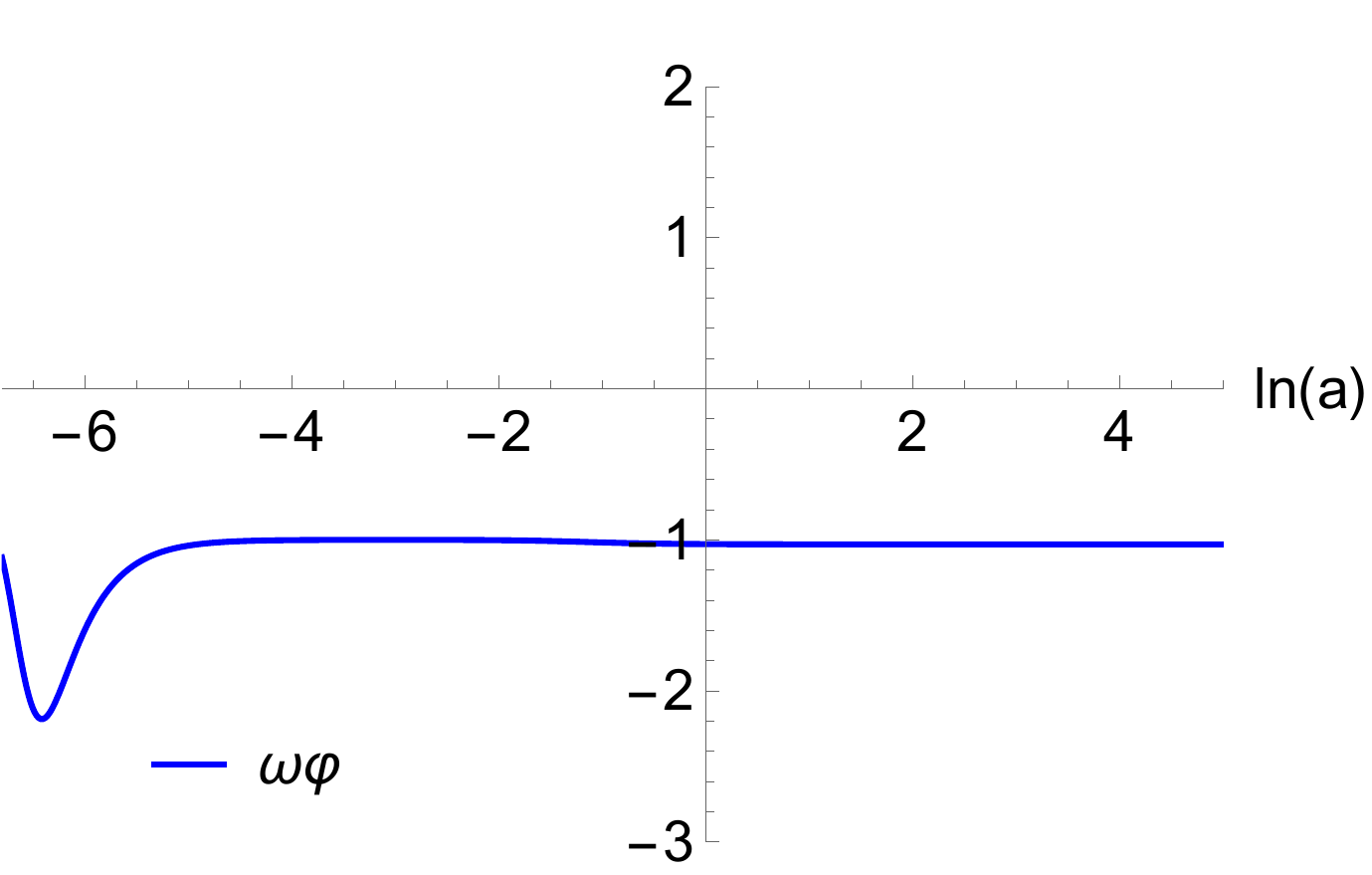}
    \caption{C = 0.001}
    \label{fig:wphi_c_0.001}
  \end{subfigure}
  \begin{subfigure}[b]{0.49\linewidth}
    \includegraphics[width=\linewidth]{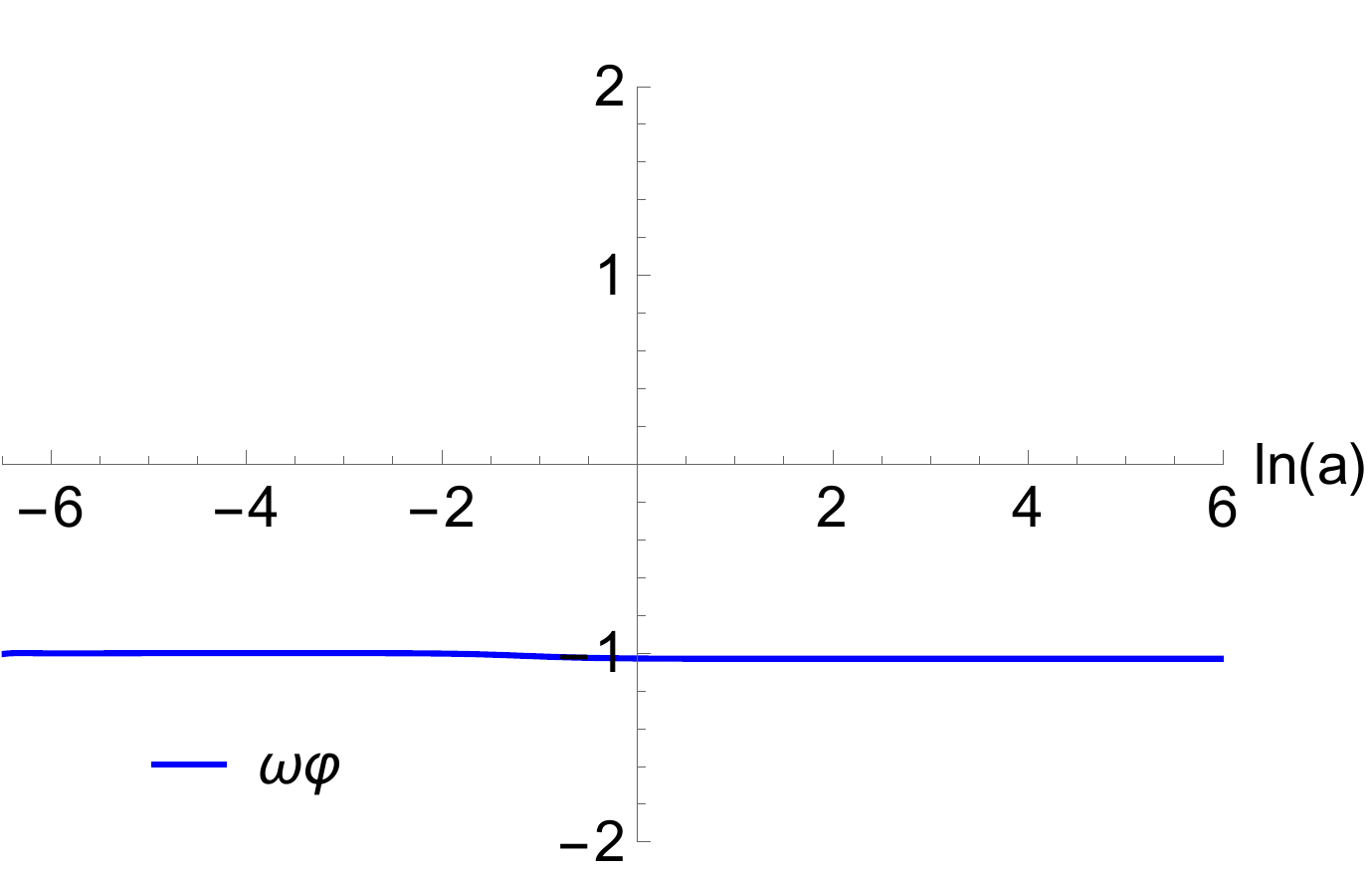}
    \caption{C = 0.0001}
    \label{fig:wphi_c_0.0001}
  \end{subfigure}
  \caption{The qualitative behaviour of the DE equation of state ($\omega_\phi$) parameter for different values of coupling constant(C) .}
  \label{fig:EoS_phi_plot}
\end{figure}

\begin{figure}[!ht]
  \centering
  \begin{subfigure}[b]{0.49\linewidth}
    \includegraphics[width=\linewidth]{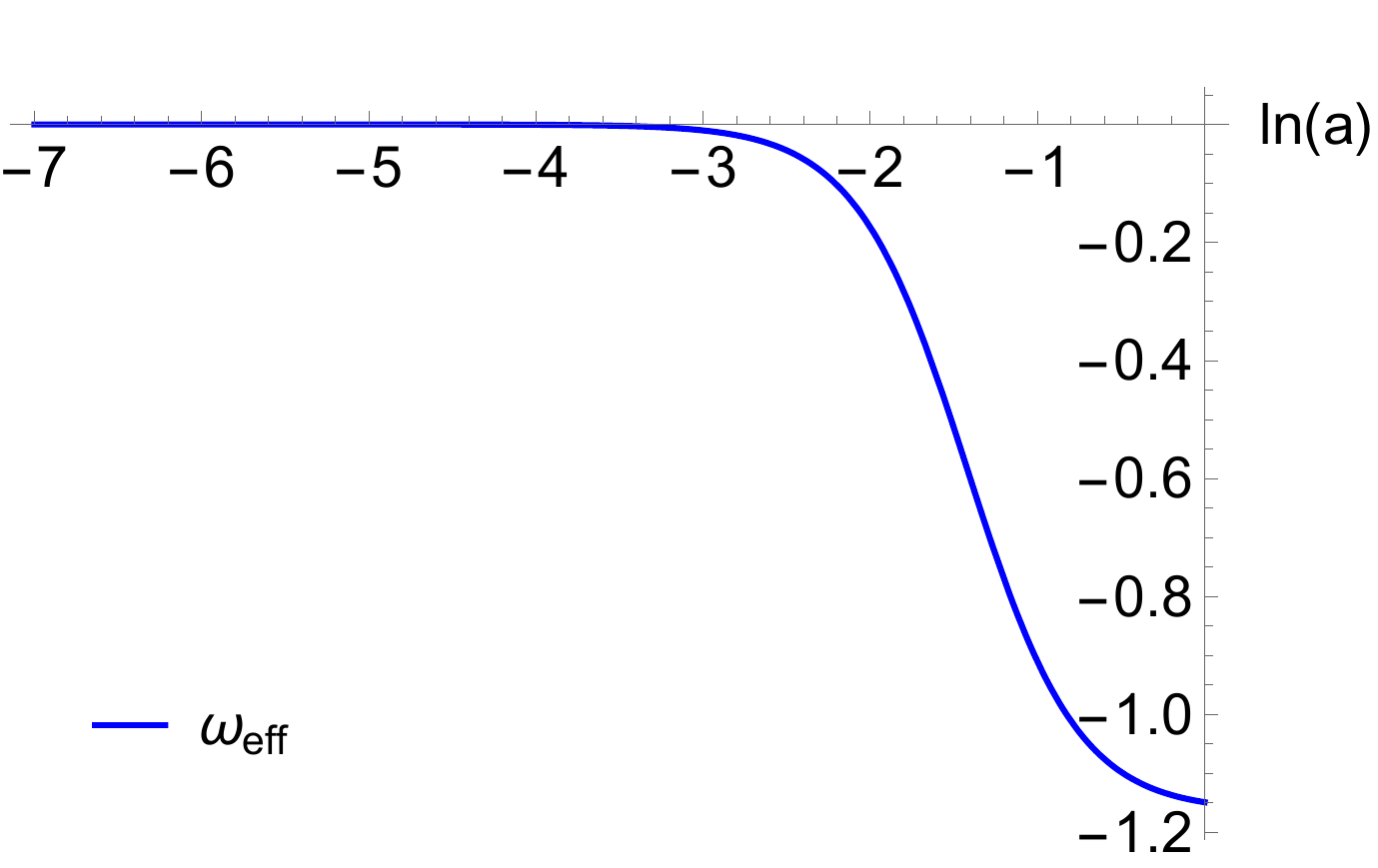}
    \caption{$\omega_{eff}$}
    \label{fig:EoS plot_eff}
  \end{subfigure}  
  \begin{subfigure}[b]{0.47\linewidth}
    \includegraphics[width=\linewidth]{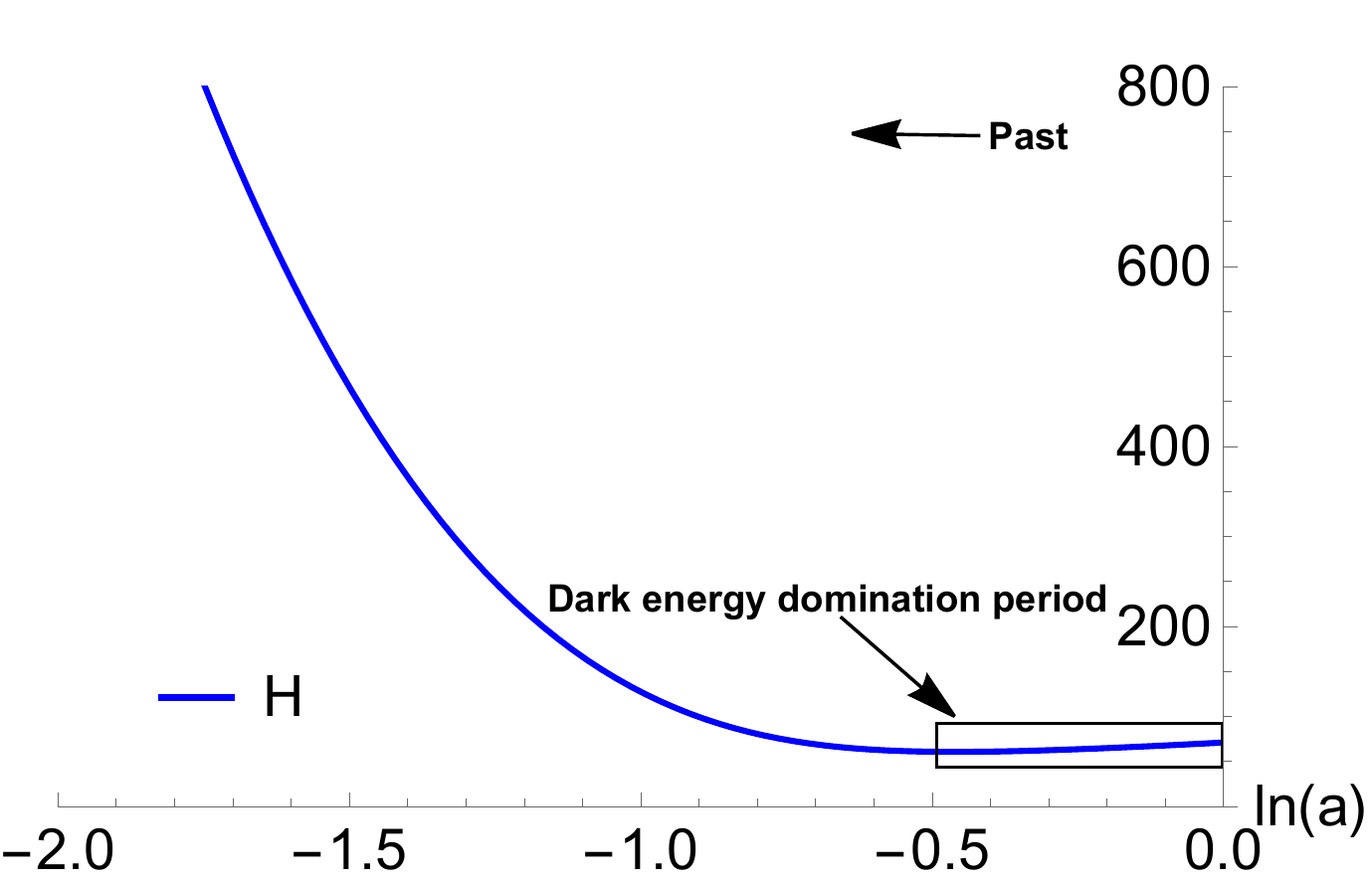}
    \caption{H[km s$^{-1}$Mpc$^{-1}$]}
    \label{fig:hz_plot}
  \end{subfigure}  
  \caption{The evolution of the effective equation of state parameter ($\omega_{eff}$) and the Hubble parameter (H) as a function of N = ln(a) for an interacting DE-DM model..}
  \label{fig:EoS_Hz_plot}
\end{figure}

The specified form of potential has already simplified the analysis of an autonomous system for coupled DE-DM model. However, it has been observed that at late times, the cosmological parameters show nearly steady behaviour with `ln$(a)$' variation (the reader can also refer \cite{PhysRevD.103.023510}). This supposition brings us a practicability to constrain the model in such a way that the term $`\left(\frac{2 \beta-k}{2}\right)\Omega_m$' in the interaction strength obtained in Eq. \eqref{eqn:35} becomes a constant (C) entity. Here, we define `C'  as a coupling constant. For further analysis of the phase-space portrait, the choice of the potential and other parameter values are as mentioned in Eq. \eqref{eqn:56}. The physical motivation behind choosing such set of parameters is that they facilitate the universe's current cosmological evolution.
\begin{align}
    V(\phi) &=V_0 e^{\alpha (\phi)},&
    B(\phi) &= sin^2(\phi),&
    k(\phi) &= -2cot(\phi),&
    \alpha(\phi) &=c' \phi, &
    \beta &= - \frac{c'}{\alpha}, \label{eqn:56}
\end{align}
\begin{figure}[!ht]
  \centering
  \begin{subfigure}[b]{0.49\linewidth}
    \includegraphics[width=\linewidth]{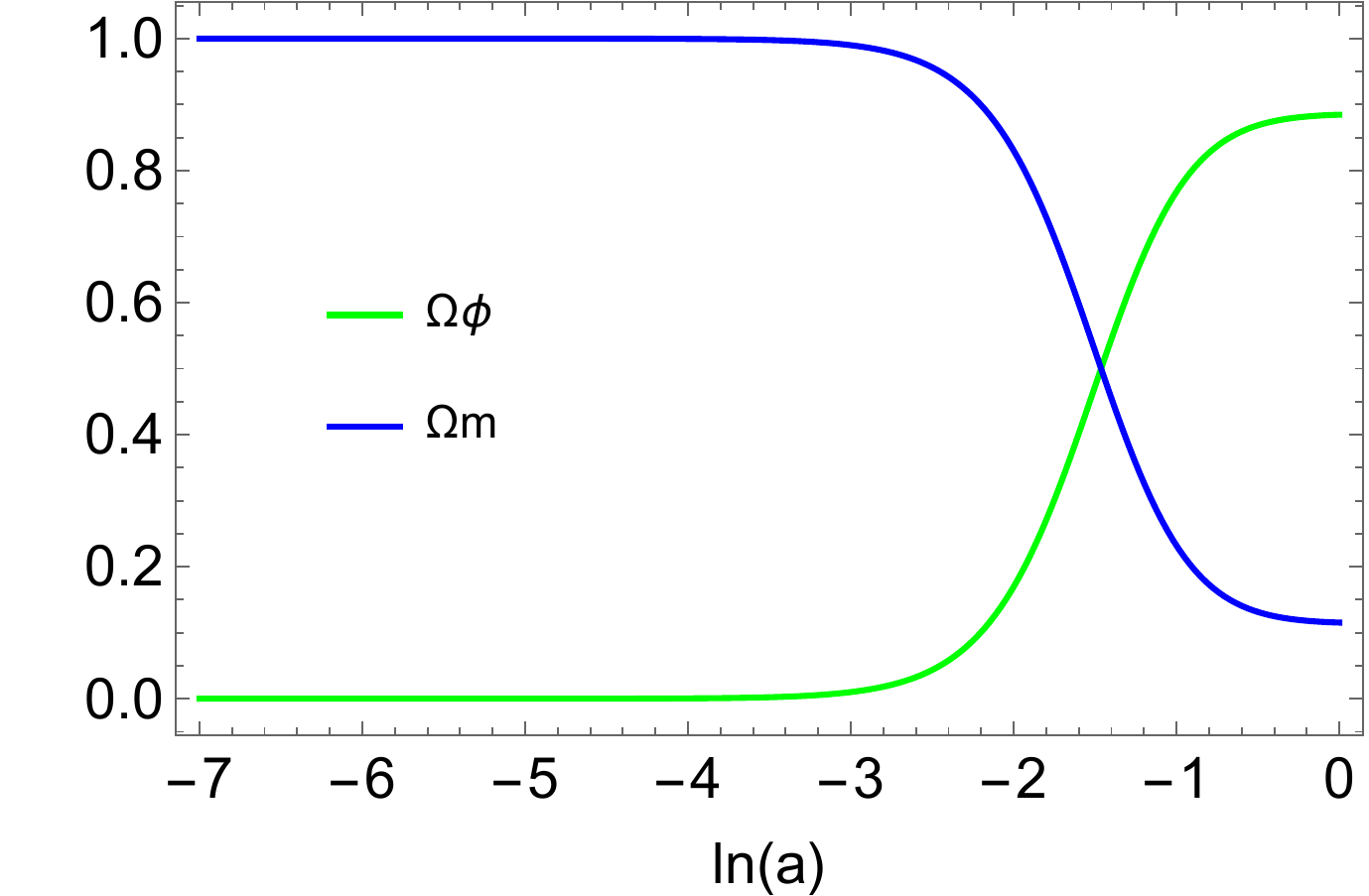}
     \caption{C = 0}
     \label{fig:density plot1}
  \end{subfigure}
  \begin{subfigure}[b]{0.49\linewidth}
    \includegraphics[width=\linewidth]{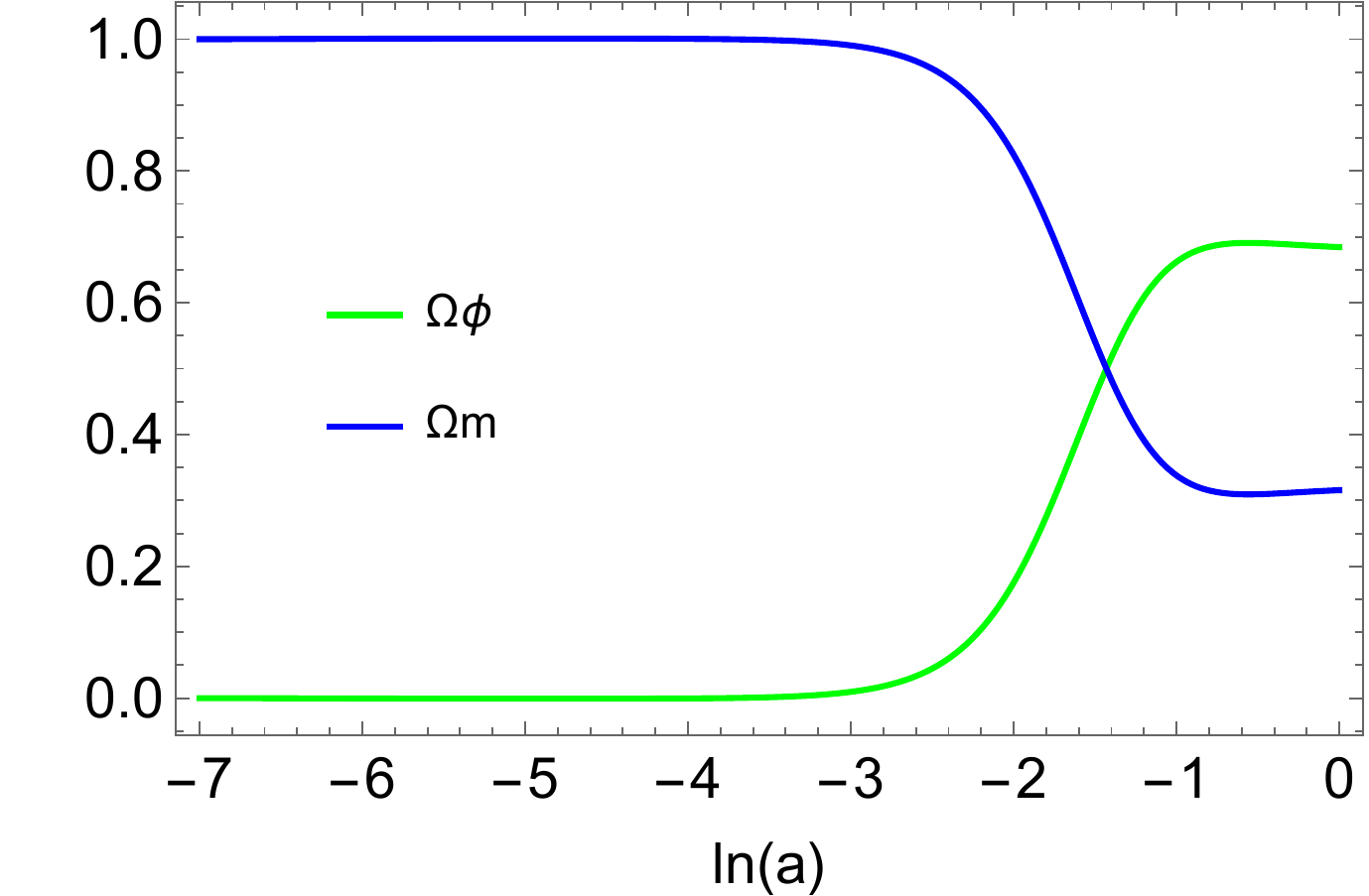}
    \caption{C = 0.04}
    \label{fig:density plot2}
  \end{subfigure}  
  \begin{subfigure}[b]{0.49\linewidth}
    \includegraphics[width=\linewidth]{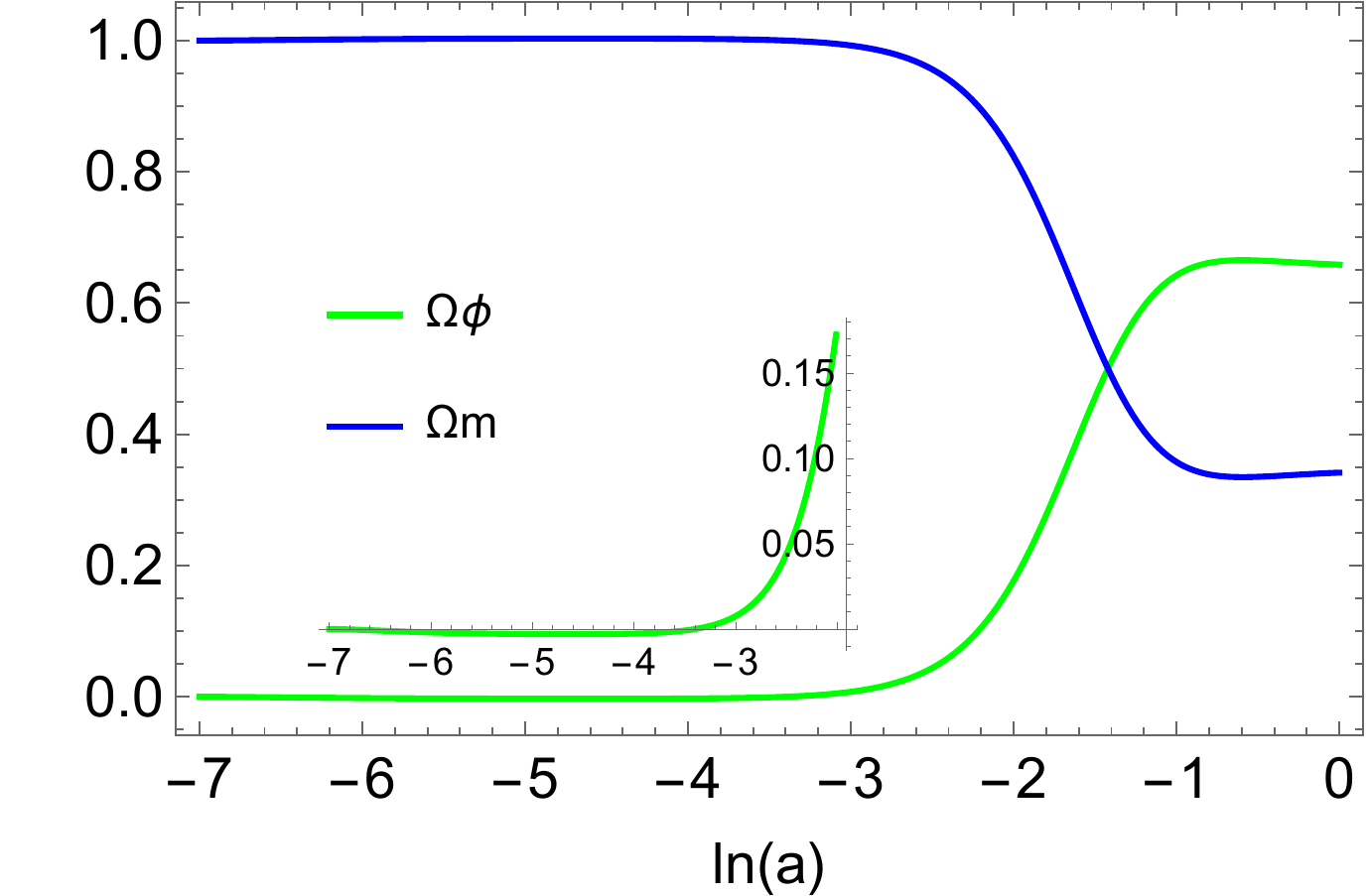}
    \caption{C = 0.07}
    \label{fig:density plot3}
  \end{subfigure}
  \begin{subfigure}[b]{0.49\linewidth}
    \includegraphics[width=\linewidth]{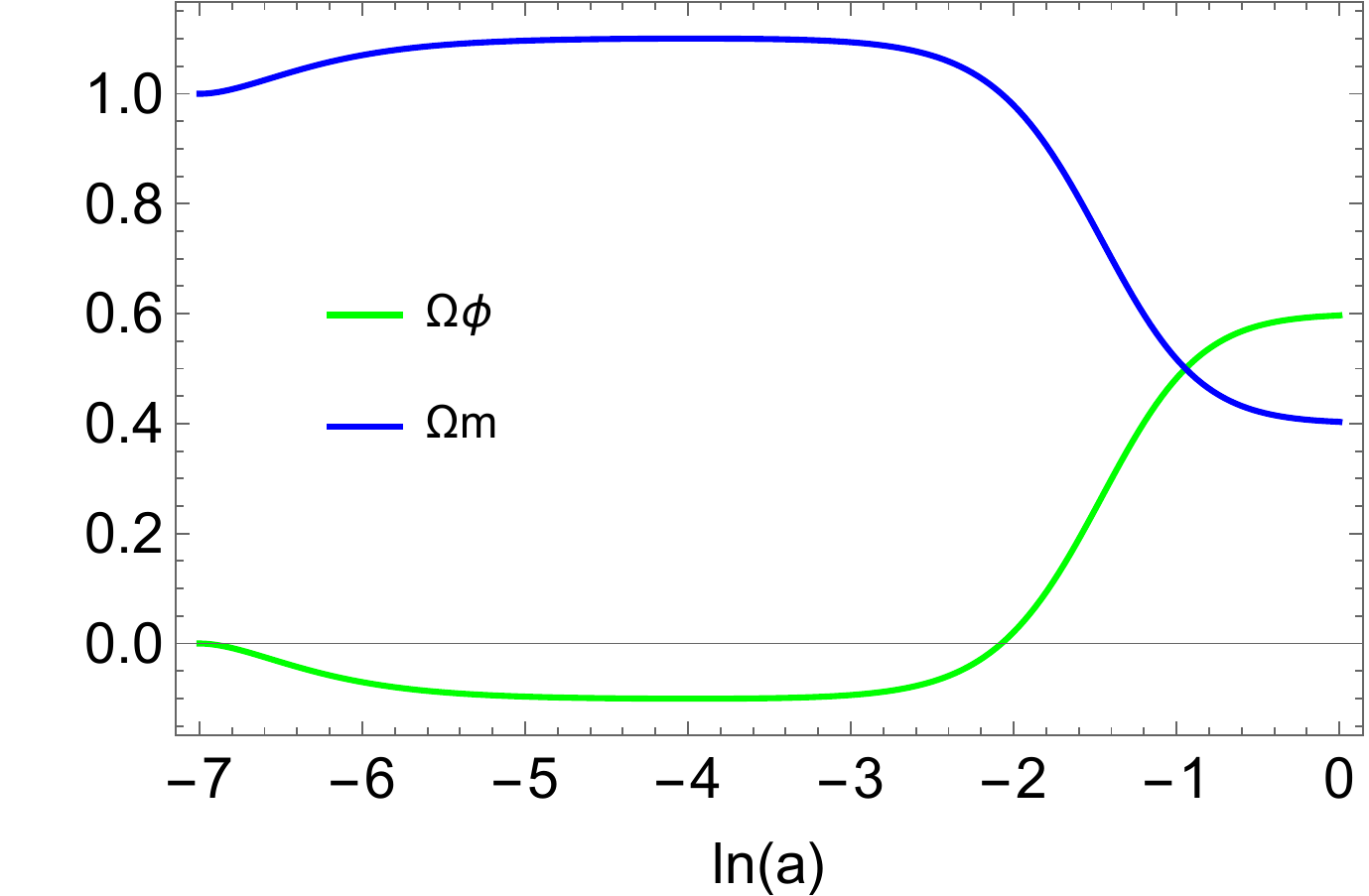}
    \caption{C = 0.4}
    \label{fig:density plot4}
  \end{subfigure}  
  \caption{The relative density parameter of DE ($\Omega_\phi$) and DM ($\Omega_m$) with varying N=ln(a) for various values of coupling constant(C).}
  \label{fig:1}
\end{figure}
where `$\phi$' varies between 0 and $\pi$. We show the cosmological evolution for positive values of coupling constant(C). Figure \ref{fig:1} represents the plots for the evolution of the relative density parameter of DE and DM as a function of `N' for different values of coupling constant. The plots clearly show the DM dominated universe over a long period of time in the past and then evolve to the DE field($\phi$) dominated universe.
Figure \ref{fig:density plot1} portrays uncoupled case and the behaviour becomes analogous to the canonical scalar field. Figure \ref{fig:density plot2} portrays the behaviour for the coupling constant $`$C' = 0.04. For such smaller coupling strengths, we get a completely viable background cosmological model. Figure \ref{fig:density plot3} indicates the onset of negative dark energy density parameter for $`$C' = 0.07 which can clearly be seen in figure \ref{fig:density plot4} for further higher values of $`$C'. We found that for coupling strength $`$C' > 0.07, the DE density parameter violates the WEC in the past thus, setting an upper bound of $`$C' = 0.07 for $\lambda$ = 1.
%(as mentioned in the section \ref{4}). 
%It must be taken into account that this violation is due to the presence of large scalar field coupling
%and hence such violation   % which quite put an upper bound on $`$C' values.
The other allowed values of $\lambda$ would make no difference on the upper limit value.
%For smaller coupling strength $`$C', the WEC holds perfectly (Fig. \ref{fig:density plot2}, Fig. \ref{fig:density plot3}) and one can have a completely viable background cosmological model. 
Thus, looking at the results we emphasize that the said violation could be avoided and the current background dynamics obeying the energy condition (WEC) can be achieved by pertaining to the small $`$C' values. Hence, we follow the smaller couplings to avoid any other such ambiguities.  For coupling constant(C) = 0.04, figure \ref{fig:density plot2} mirrors the current observational measurements of DE and DM density in the present-day observable universe\footnote{The dominance of the dark energy over the dark matter takes place at later time if we further lower the $``\lambda$" value.}. This may equip us with derived empirical results of the coincidence problem. Also, we have observed that for allowed $`$C' values,  the results are compatible with the current evolution of the dark energy EoS parameter, $``\omega_\phi"$ (Fig. \ref{fig:EoS_phi_plot}), the effective EoS parameter, $``\omega_{eff}"$ (Fig. \ref{fig:EoS plot_eff}) and Hubble parameter, $``$H" (Fig. \ref{fig:hz_plot}), respectively (refer \cite{Freedman:2019jwv,Riess:2019cxk,Chen:2019ejq,Wong:2019kwg,Anagnostopoulos:2019myt,Yang:2019jwn,Shafer:2013pxa,Riess_2011} and references therein.). 
\begin{figure}[ht]
    \centering
    \includegraphics[width=0.6\textwidth]{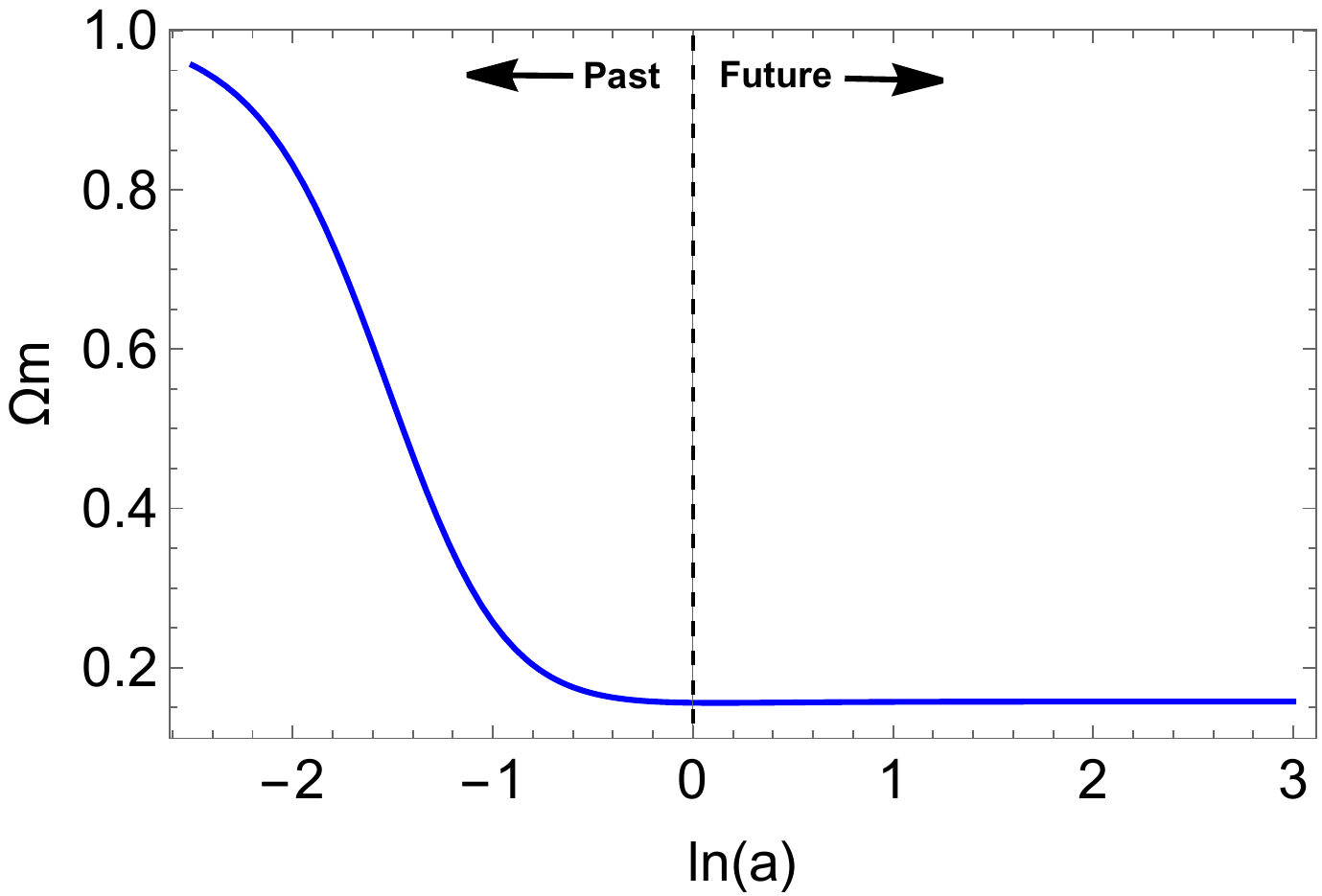}
    \caption{Late time evolution of DM energy density for an interacting DE-DM model. }
    \label{fig:2}
\end{figure}
\par
The observation of figure \ref{fig:2} suggests that the density parameter $\Omega_{m}$ changes steadily at late times. Also, to ease the analysis of the system, we select some particular values of the parameter `k' along with `$\beta$' where the coupling term $\frac{(2\beta-k)}{2}\Omega_m$ takes values as C = 0, C = 0.005, C = 0.04. We are interested in positive values of coupling strength\footnote{The negative coupling constant produces approximately same results (the reader can refer \cite{PhysRevD.103.023510})} and hence the analysis has been performed only for those values.
\begin{figure}[h!]
  \centering
  \begin{subfigure}[b]{0.45\linewidth}
    \includegraphics[width=\linewidth]{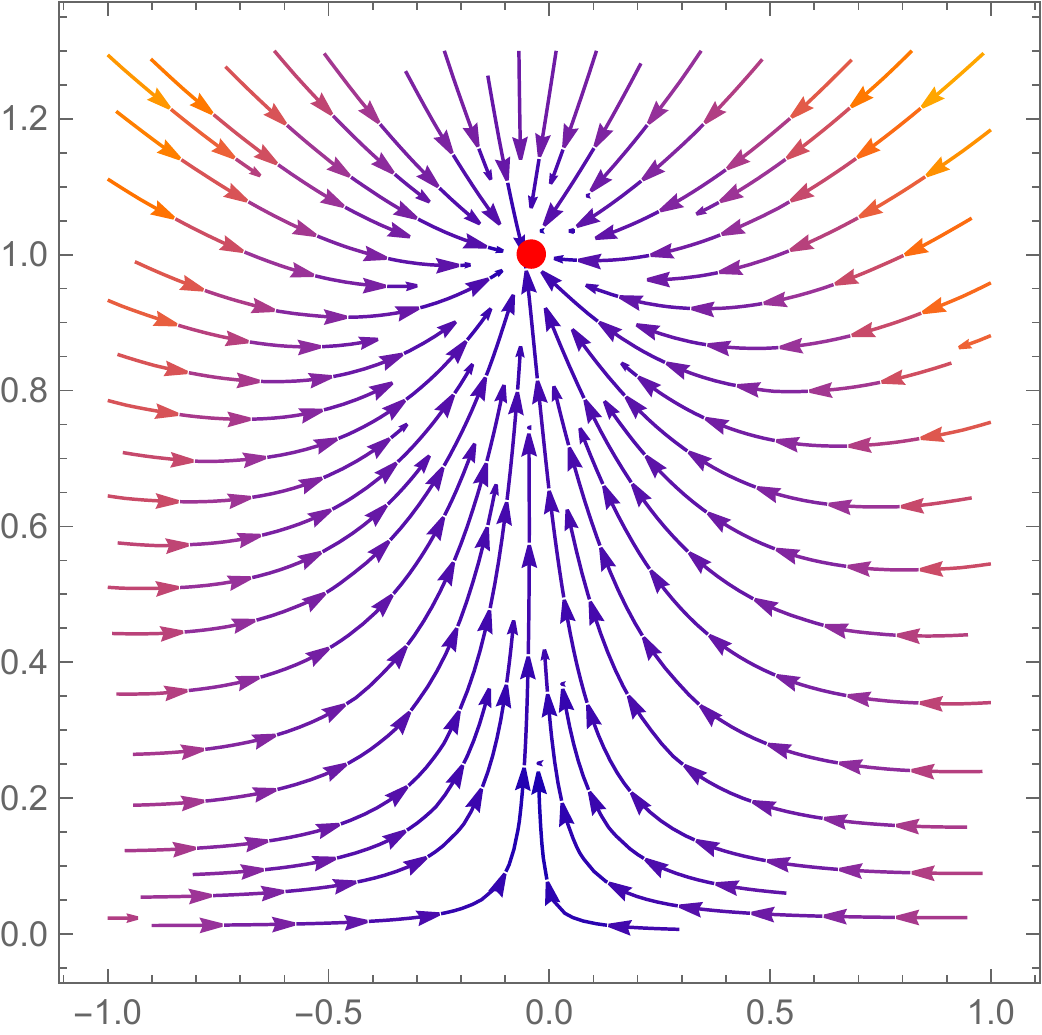}
     \caption{$\lambda=0.1$}
  \end{subfigure}
  \begin{subfigure}[b]{0.45\linewidth}
    \includegraphics[width=\linewidth]{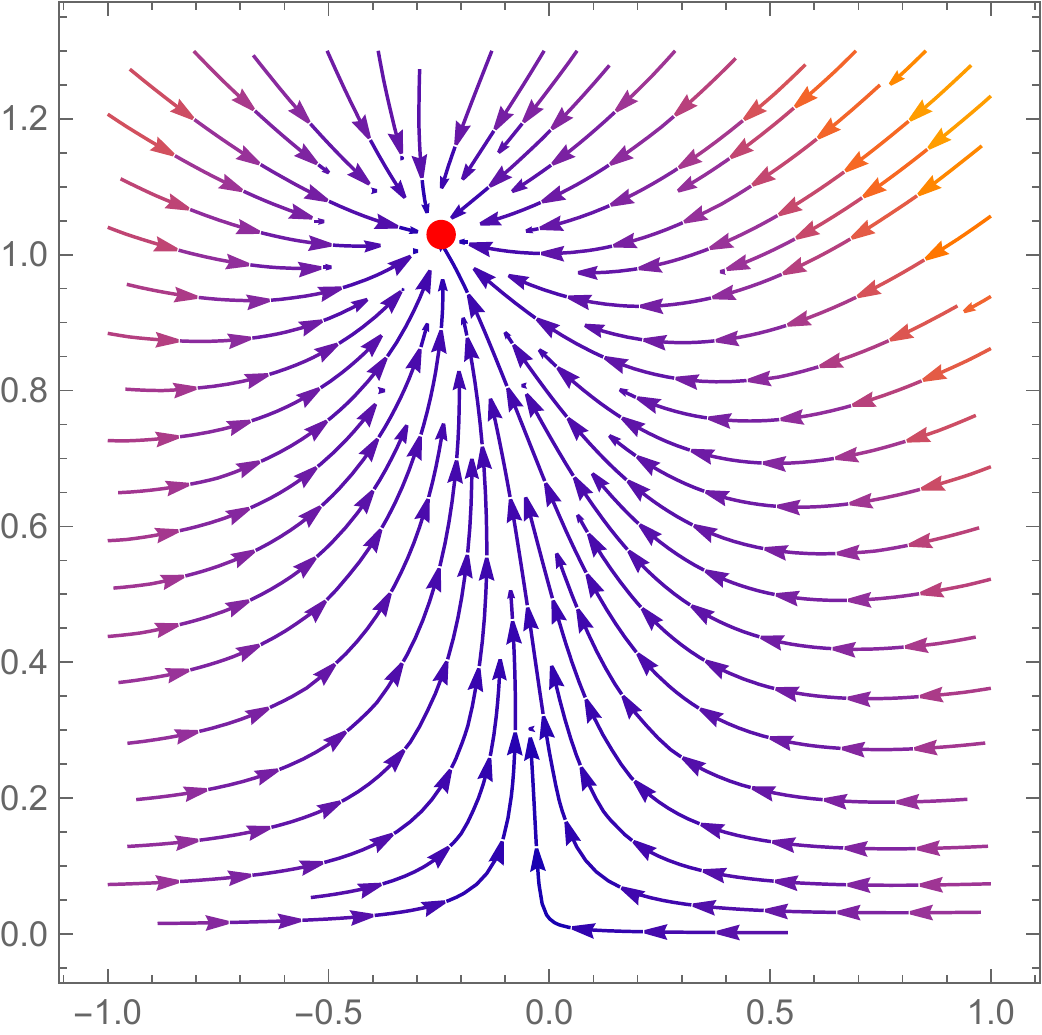}
    \caption{$\lambda=0.6$}
  \end{subfigure}
    \begin{subfigure}[b]{0.45\linewidth}
    \includegraphics[width=\linewidth]{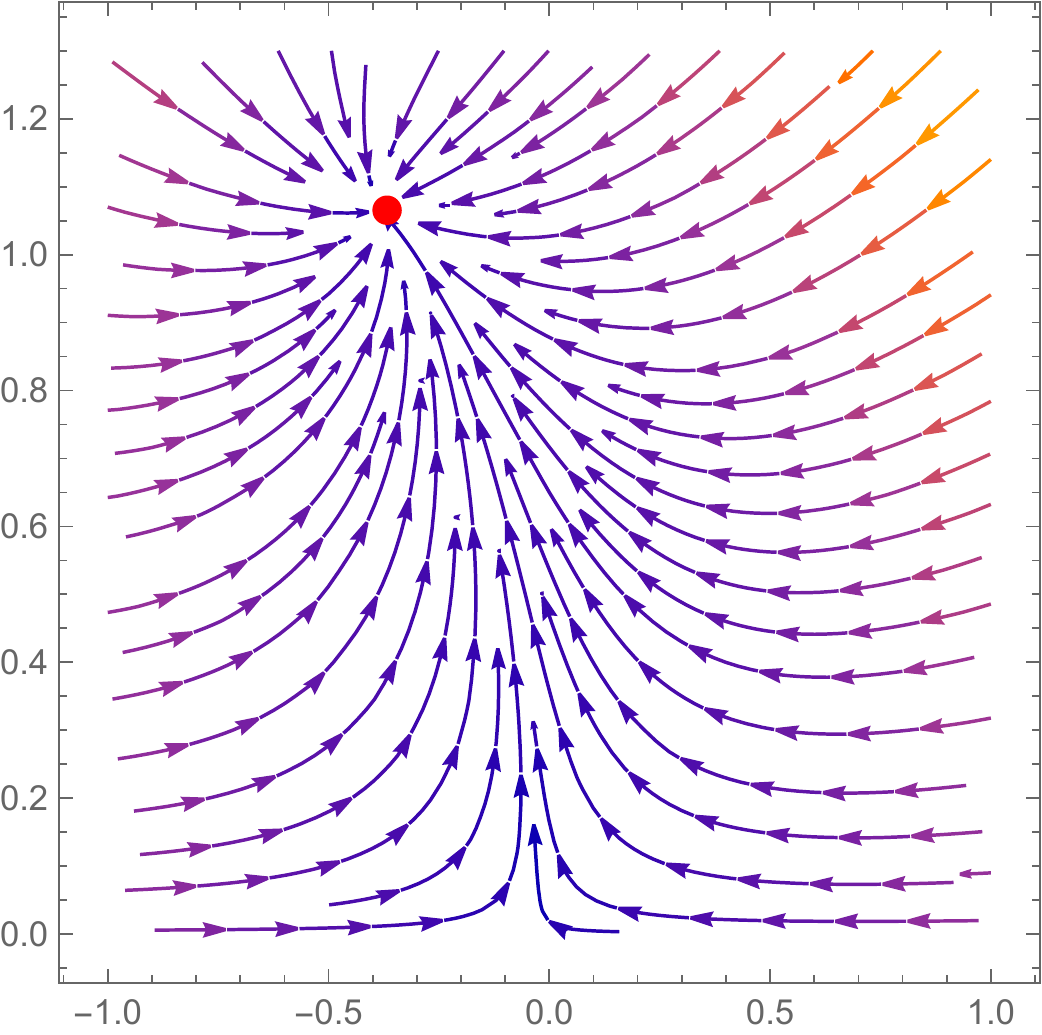}
    \caption{$\lambda=0.9$}
  \end{subfigure}
  \caption{The phase space portraits of the dynamical system for an equilibrium point D for $\lambda=0.1$, $\lambda=0.6$ and $\lambda=0.9$ with a coupling constant ‘C’= 0.04. The horizontal axis represents variable `x' and vertical axis represents variable `y'.}
  \label{fig:phase portrait plot}
\end{figure}
\par
Our focus on the late-time cosmological behavior of the universe makes this feature agreeable to examining the time evolution of variables x and y in figure \ref{fig:phase portrait plot}. We observe the dark energy dominated solution for a coupled DE-DM model for critical point D represented by a red dot in figure \ref{fig:phase portrait plot}. All the trajectories moving towards point D signals the stable attractor nature of the critical point and the figure exhibits an everlasting late-time accelerated nature of the universe for $\lambda^2 >-2$. The phase space portraits are shown for different values of $\lambda$ and thus, we reveal that regardless of the initial conditions the qualitative behavior of the universe at the background level nearly remains the same. It can also be spotted that all the trajectories are getting attracted towards a single trajectory and not crossing that specific one. These dynamics urge us to proclaim that trajectory itself, in this incident, is acting as an attractor. We call them a late-time acceleration trajectory similar to an inflationary trajectory (separatrix) in an inflationary scenario.
\par
Figure \ref{fig:EoS_phi_plot} presents the cosmological evolution of DE EoS parameter. It can be seen that in recent times, the dynamics of the evolution are dominated by the DE field $``\phi"$. We extrapolate that the behaviour corresponds to phantom regime. On the other hand, the dynamics of the effective EoS parameter, $\omega_{eff}$ tracks the quintessence dominated nature $``-1<\omega_{eff}<-\frac{1}{3}"$ to phantom dominated nature $``\omega_{eff}<-1"$ in the present time (Fig. \ref{fig:EoS plot_eff}). Hence, the analysis clearly signifies the present time accelerated expansion of the universe.
\begin{figure}[!ht]
  \centering
  \begin{subfigure}[b]{0.48\linewidth}
    \includegraphics[width=\linewidth]{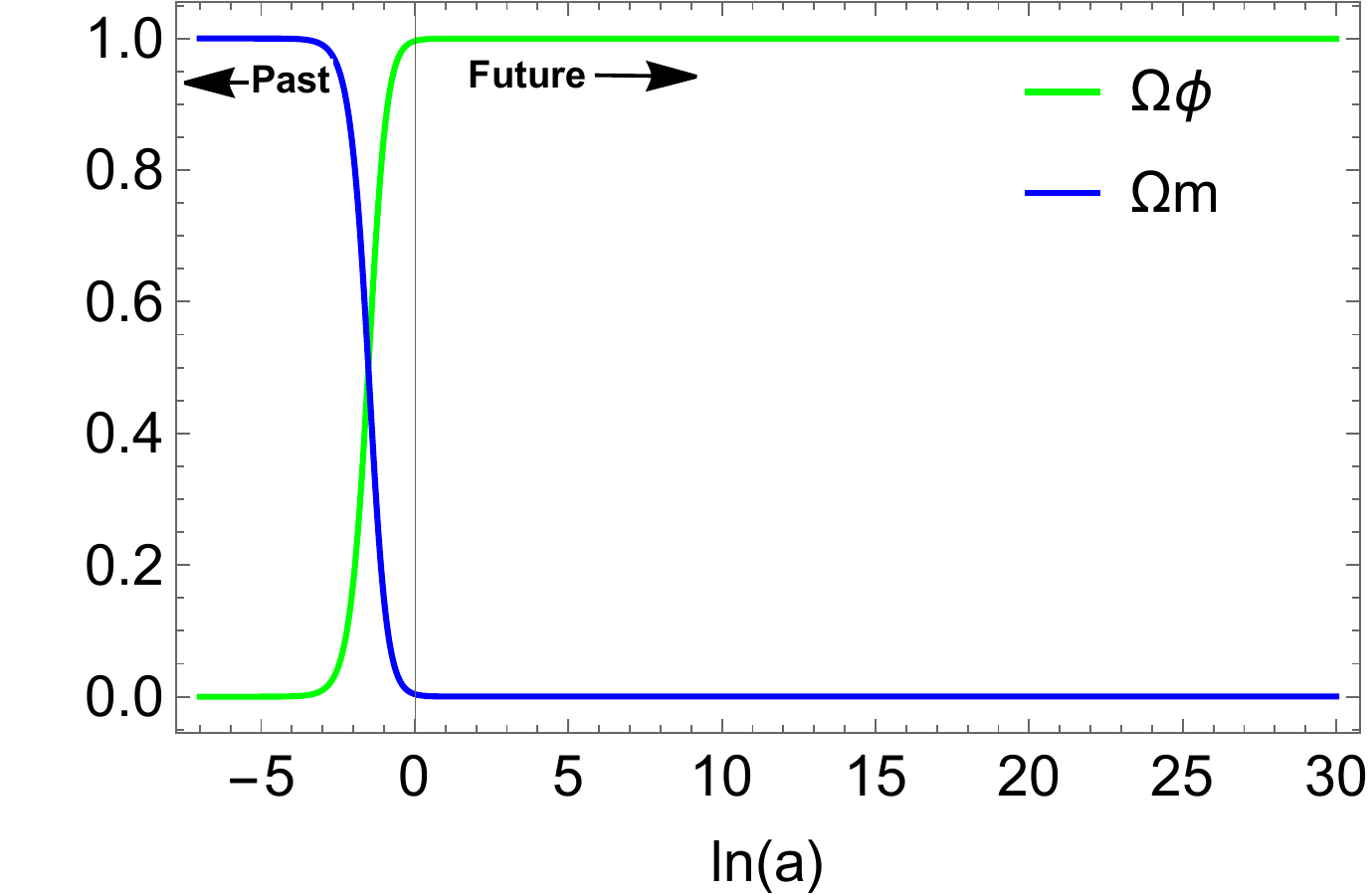}
    \caption{C = 0.009}
    \label{fig:no_singularity_a}
  \end{subfigure}
    \begin{subfigure}[b]{0.48\linewidth}
    \includegraphics[width=\linewidth]{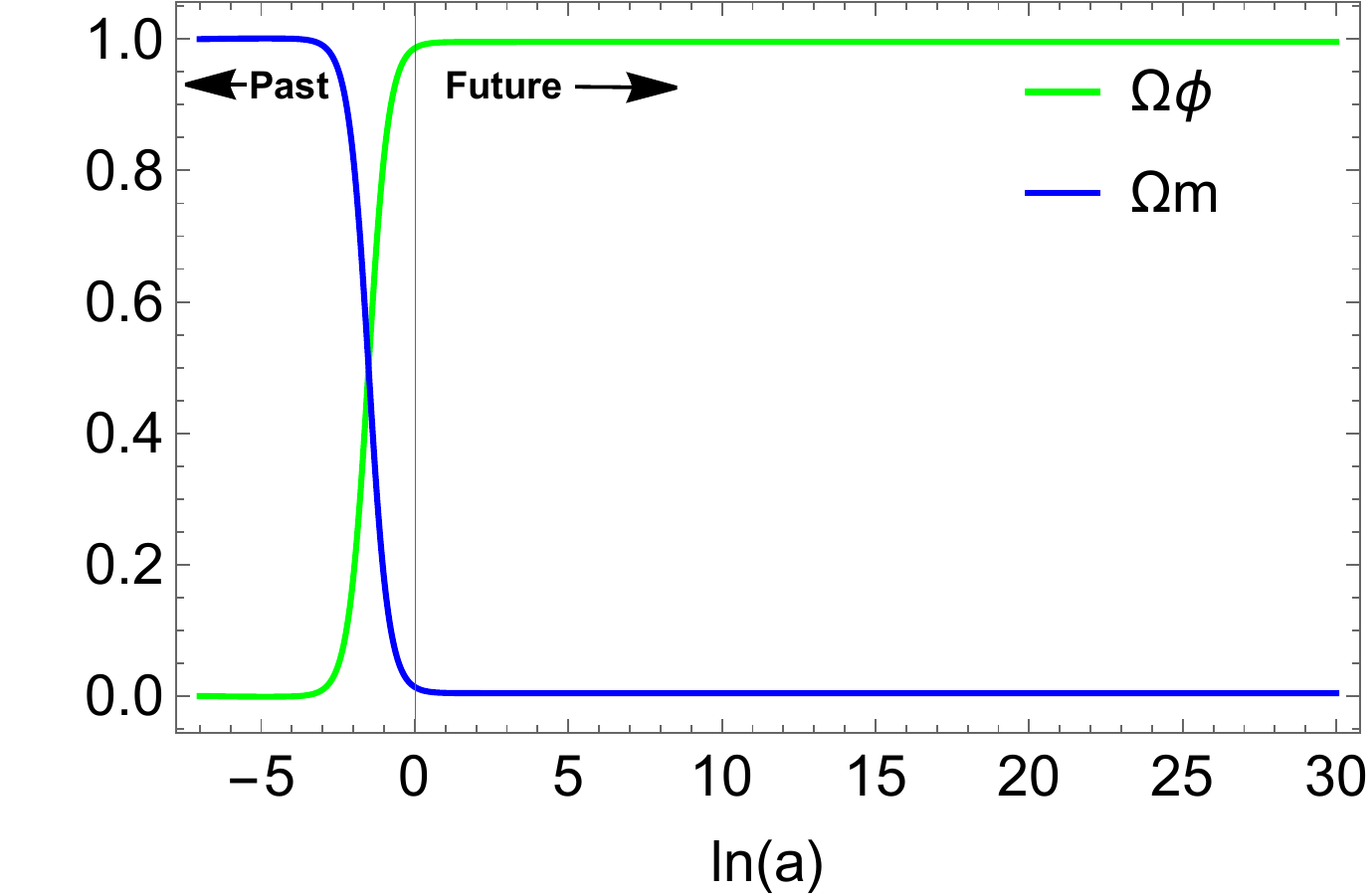}
    \caption{C = 0.04}
    \label{fig:no_singularity_b}
  \end{subfigure}
  \caption{The variation of density parameter $\Omega_\phi$ and $\Omega_m$ for couplings C = 0.009 and C = 0.04 exhibiting no appearance of future singularity.}
  \label{fig:no singularity plot}
\end{figure}

\par
Furthermore, from the figure \ref{fig:wphi_c_0.8} it can be seen that $\omega_\phi$ attains larger negative values at the onset of the dark energy component domination. This feature results from the coupling effect on the $\omega_\phi$ behaviour. If we reduce the coupling, the $\omega_\phi$ becomes less negative (see Fig. \ref{fig:wphi_c_0.001} \& Fig. \ref{fig:wphi_c_0.0001}) as permitted by the observations \cite{PhysRevD.68.023522,Tripathi:2016slv,BOSS:2016wmc,Nesseris:2006er}. (Also, It is found that small coupling $``$C" have minimal effect on the evolution of density parameters (refer Fig. \ref{fig:no_singularity_a})).  And if we further increase the $``$C"\footnote{Here, we found that dark energy do not overtake the dark matter for large $``$C" values which is not physically acceptable scenario. Therefore, we restrict the coupling to be minimal to achieve the necessary dynamics. Nevertheless, these dynamics again depend upon the constraint put on the parameter $``\lambda"$.}, the feature remains the same (with large negative value) but gets shift in the future (Fig. \ref{fig:wphi_c_1.2}). With this knowledge in mind, we find from the figure \ref{fig:no singularity plot} that there is no appearance of singularity even in the far future regardless of the size of the coupling. All these findings are based upon the consideration that whether introduction of interaction and by how much amount causes any type of singularity or not. Here, we have observed that irrespective of the coupling considered, the behaviour of $\omega_\phi$ stabilizes at $\sim$ -1 at present time (See Fig. \ref{fig:EoS_phi_plot}). Thus, in our investigation we find with the constraints on this EoS parameter that, this particular model of interaction between dark matter and dark energy gives rise to the accelerated expansion of the universe at late times and can also help to avoid the possibility of big rip singularity (if there is any in the future) since EoS parameter always tends towards -1 ($\omega_\phi \to$ -1) in the present time and even in the future (Fig. \ref{fig:EoS_phi_plot}), giving us an attractor de Sitter like solution on the large scale.\footnote{This has already been investigated in this study.}
\par
To summarize, the constraints on the proposed DE-DM interaction model restrict the potential and the values of other parameters facilitating the currently observed evolution of the universe, as shown in \cite{Mandal:2021ekc,PhysRevD.103.123517,Dutta:2018xcz}. The dynamical stability analysis and the analysis from figures \ref{fig:1}-\ref{fig:EoS_Hz_plot} suggest that with the appropriate choice of parameters $\beta$ and $\lambda$, the universe exhibits DM dominated nature long enough to let the structure formations to happen. Later, it evolves to DE (field $`\phi$') dominated universe causing an accelerated expansion in agreement with the current cosmological observations.

\section{Conclusions}  \label{6}
In this article, we have put forward the representation of the two-field dark sector model to describe the accelerated expanding universe and also to test the feasibility of the model in the background cosmology under the convenience of chiral cosmology. We set up a general formulation method to study 2-field chiral model. Further, approaching this point of view, to study the dark sector fields' evolution. We proposed an interacting canonical and non-canonical scalar field model termed as coupled DE-DM model. We presented that there is a resemblance between the field theory approach and phenomenological fluid approach to study the DE-DM interaction model. This method makes all the cosmological equations appear in terms of DM fluid and DE scalar field $\phi$, creating possible scenarios to formulate the dynamical system of equations effortlessly. In our case, we have opted not to surplus the system of equations signaling the radiation-dominated era and thus, focused on the dark matter and dark energy dominated period of the universe. We chose this scenario since we are interested in studying the late time cosmological behaviour. 
\par
Later, we conducted a detailed fixed point analysis and stability analysis of the dynamical system set up for an interacting DE-DM model. The chosen exponential potential of the form $V(\phi)=V_0 e^{\alpha(\phi)}$ and parameters constraints reduced the dimensions of the system to make further analysis even more comfortable. The critical points obtained are of non-hyperbolic nature. However, the linear stability theory can still be employed to decide the nature of the points A, B, and D but vaguely. On the other hand, point C contains cosmological parameters of higher power terms and the stability analysis is found to be convoluted.
\par
For the critical point A, we have gained parameter restrictions to be $\Gamma_k <1$ indicates the saddle nature of the point but $\Gamma_k >1$ leaves the stability undetermined. Also, the vanishing $\omega_{eff}$ makes this point unsuitable for acceleration. The critical point B has been analysed with the employment of the centre manifold theory and its stability features have correspondingly been portrayed in figure \ref{fig:criticalpointB_vectorfield}. Point B describes both quintessence and phantom DE dominated accelerated expansion as seen from table \ref{tab:table1}. The cosmological evolution corresponding to point C describes the scaling solution and also describes an accelerated era of the universe for parameters with $-2 < \frac{\lambda}{ \beta}<1$ constraint. It is to be noted that for vanishing $\frac{\lambda}{ \beta}$, point C nurtures the de Sitter universe. The point gains the cosmological importance since the late time accelerated scaling solution can alleviate the coincidence problem as well with few constraints on parameters. The figure \ref{fig:region plot} facilitate us with the allowed range of parameters to understand the stable and unstable behaviour of point C accordingly. The non-hyperbolic critical point D also exhibits an accelerating universe under $\lambda^2 >-2$ parameter constraint. Point D has been analysed with the application of the centre manifold theory and its stability features are corresponding as shown in table \ref{tab:table4}. For specific values of parameter $\lambda$, the point is a stable attractor (Fig. \ref{fig:phase portrait plot}) and describes the cosmological constant or phantom field or quintessence field dominated universe.
\par
To complement the above analysis, in our work, we have also discussed the graphical presentation of the dynamical system. In figure \ref{fig:1} we studied the evolution of density parameters of dark matter and dark energy. We set our analysis to be short of radiation component supposing that the universe has encountered a radiation dominated era long before it evolved to the time period dominated by the dark matter which is in late time follows the dark energy dominated behaviour as it is seen from figure \ref{fig:1}. Figure \ref{fig:2} shows the era where the evolution of dark matter energy density becomes almost constant after it went through an era that is long enough for all kinds of structure formation to take place. Later we discussed the phase-space portrait of the system (Fig. \ref{fig:phase portrait plot}) picturing the stable attractor D determining the late time DE dominated accelerated evolution of the universe. The evolution of the equation of state parameters in figures \ref{fig:EoS_phi_plot}-\ref{fig:EoS plot_eff} also shows cosmologically relevant behaviour complimenting the aforesaid analysis of the dynamical system. This investigation also implies that the occurrence of future singularity problem is absent and even if there is any it can be cured.
\par
The above analysis suggests that Chiral cosmological theory can be used to sense the cosmologically viable solutions in the realm of background dynamics, where we notice a sufficiently long, extended era of the matter-dominated universe to an era of an accelerated expanding universe in the present time. Our next plan is to employ the observational data analysis methods and be more precise about the parameter values encountered in the current study. In the future, we also plan to investigate other relevant cosmological models \cite{PhysRevD.87.084031,Sa:2020qfd,Harko_2020,Das:2020kff} under the Chiral theory of cosmology and also plan to analyze them by bringing in the cosmological experimental framework. 

\section*{Acknowledgement}
This work is partially supported by DST (Govt. of India) Grant No. SERB/PHY/2021057.
\section*{Appendix}
Since the output corresponding to eigenvalues of critical point C is too big to handle and it is a tedious task to display them on the full scale, we struck out a large number of terms and simply intended to disclose the mannerism in which it appears during the course of action which makes them inconvenient to deploy for the analysis of stability behaviour of the corresponding fixed point.
\begin{multline*}
    \Bigg[0, \frac{1}{(\beta-\lambda)^4} \sqrt{(2592 \beta^9-1728\beta^{11})  (\beta-\lambda) -5760 \beta^{11}  (\beta-\lambda) \lambda^2 +...  +(96 \beta^{9} \lambda+...} ,\\[1.0pt]
    \frac{1}{(\beta-\lambda)^4} \sqrt{-1728\beta^{11} (\beta-\lambda) +(20736 \beta^{8}  \lambda +72576 \beta^7 \lambda^2 )(\beta-\lambda)+... +(72 \beta^{6} \lambda^{2}+...} ,\\[1.0pt]
    \frac{1}{(\beta-\lambda)^4} \sqrt{-1728\beta^{11} (\beta-\lambda) +...-10368  (\beta-\lambda)\beta^{4} \lambda^7   +(96 \beta^{9} \lambda+.....
    -1296 \beta \lambda^5)+...}
    \Bigg].
\end{multline*}

\begin{center}
 \rule{4in}{0.1pt}\\
 \vspace{-13pt}\rule{3in}{0.1pt}\\
 \vspace{-13pt}\rule{2in}{0.1pt}\\
  %\medskip
\end{center}

\begin{spacing}{0.0}
\bibliography{DE_DM_int}
\bibliographystyle{ieeetr}
\end{spacing}

%\begin{multicols}{2} %split the mulitcol to start the reference entirely after the text
%\printbibliography[heading=none]
%\end{multicols}

\end{document}